%% file: gravi-agn_ngc3783_astro-ph.tex
\documentclass[longauth,]{aa}  
\usepackage{graphicx}
\usepackage{array}
\usepackage{txfonts}
\usepackage{hyperref}
\hypersetup{colorlinks=true, citecolor=blue, linkcolor=blue}
\usepackage{subfig}
\usepackage[dvipsnames]{xcolor}

\usepackage{gensymb}
\newcommand{\micron}{{\mbox{$\mu$m}}}
\newcommand{\brg}{Br$\gamma$}
\newcommand{\caviii}{[\ion{Ca}{viii}]}
\newcommand{\sivi}{[\ion{Si}{vi}]}
\newcommand{\mbh}{{\mbox{$M_\mathrm{BH}$}}}
\newcommand{\uas}{{\mbox{$\mu$as}}}

\begin{document}

   \title{The central parsec of NGC~3783:  a rotating broad emission line region, asymmetric hot dust structure, and compact coronal line region}
   \titlerunning{The central parsec of NGC~3783}
   \authorrunning{GRAVITY Collaboration}
\author{GRAVITY Collaboration\thanks{GRAVITY is developed
    in a collaboration by the Max Planck Institute for
    extraterrestrial Physics, LESIA of Observatoire de Paris/Universit\'e PSL/CNRS/Sorbonne Universit\'e/Universit\'e de Paris and IPAG of Universit\'e Grenoble Alpes /
    CNRS, the Max Planck Institute for Astronomy, the University of
    Cologne, the CENTRA - Centro de Astrofisica e Gravita\c c\~ao, and
    the European Southern Observatory.}:
A.~Amorim\inst{19,21}
\and M.~Baub\"ock\inst{1} 
\and W.~Brandner\inst{22} 
\and M.~Bolzer\inst{1} 
\and Y.~Cl\'enet\inst{2} 
\and R.~Davies\inst{1}
\and P.~T.~de~Zeeuw\inst{1,17} 
\and J.~Dexter\inst{24,1}
\and A.~Drescher\inst{1,27} 
\and A.~Eckart\inst{3,18} 
\and F.~Eisenhauer\inst{1} 
\and N.M.~F\"orster~Schreiber\inst{1} 
\and F.~Gao\inst{1} 
\and P.~J.~V.~Garcia\inst{15,20,21} 
\and R.~Genzel\inst{1,4} 
\and S.~Gillessen\inst{1} 
\and D.~Gratadour\inst{2,25} 
\and S.~H\"onig\inst{5}
\and D.~Kaltenbrunner\inst{1}
\and M.~Kishimoto\inst{6} 
\and S.~Lacour\inst{2,16} 
\and D.~Lutz\inst{1} 
\and F.~Millour\inst{7}  
\and H.~Netzer\inst{8} 
\and T.~Ott\inst{1} 
\and T.~Paumard\inst{2} 
\and K.~Perraut\inst{12} 
\and G.~Perrin\inst{2} 
\and B.~M.~Peterson\inst{9,10,11} 
\and P.~O.~Petrucci\inst{12} 
\and O.~Pfuhl\inst{16}
\and M.~A.~Prieto\inst{23}  
\and D.~Rouan\inst{2}
\and J.~Sanchez-Bermudez\inst{26}
\and J.~Shangguan\inst{1}
\and T.~Shimizu\inst{1}\thanks{Corresponding author: T. Shimizu (e-mail: shimizu@mpe.mpg.de)} 
\and M.~Schartmann\inst{1}
\and J.~Stadler\inst{1}
\and A.~Sternberg\inst{8,14} 
\and O.~Straub\inst{1} 
\and C.~Straubmeier\inst{3} 
\and E.~Sturm\inst{1} 
\and L.~J.~Tacconi\inst{1} 
\and K.~R.~W.~Tristram\inst{15}  
\and P.~Vermot\inst{2} 
\and S.~von~Fellenberg\inst{1}
\and I.~Waisberg\inst{13} 
\and F.~Widmann\inst{1} 
\and J.~Woillez\inst{16}}

\institute{
Max Planck Institute for Extraterrestrial Physics (MPE), Giessenbachstr.1, 85748 Garching, Germany
\and LESIA, Observatoire de Paris, Universit\'e PSL, CNRS, Sorbonne Universit\'e, Univ. Paris Diderot, Sorbonne Paris Cit\'e, 5 place Jules Janssen, 92195 Meudon, France
\and I. Institute of Physics, University of Cologne, Z\"ulpicher Stra{\ss}e 77, 50937 Cologne, Germany
\and Departments of Physics and Astronomy, Le Conte Hall, University of California, Berkeley, CA 94720, USA
\and Department of Physics and Astronomy, University of Southampton, Southampton, UK
\and Department of Physics, Kyoto Sangyo University, Kita-ku, Japan
\and Universit\'e C\^ote d'Azur, Observatoire de la C\^ote d'Azur, CNRS, Laboratoire Lagrange, Nice, France
\and School of Physics and Astronomy, Tel Aviv University, Tel Aviv 69978, Israel
\and Department of Astronomy, The Ohio State University, Columbus, OH, USA
\and Center for Cosmology and AstroParticle Physics, The Ohio State University, Columbus, OH, USA
\and Space Telescope Science Institute, Baltimore, MD, USA
\and Univ. Grenoble Alpes, CNRS, IPAG, 38000 Grenoble, France
\and Department of Particle Physics and Astrophysics, Weizmann Institute of Science, Rehovot 76100, Israel
\and Center for Computational Astrophysics, Flatiron Institute, 162 5th Ave., New York, NY 10010, USA
\and European Southern Observatory, Casilla 19001, Santiago 19, Chile
\and European Southern Observatory, Karl-Schwarzschild-Str. 2, 85748 Garching, Germany
\and Sterrewacht Leiden, Leiden University, Postbus 9513, 2300 RA Leiden, The Netherlands
\and Max Planck Institute for Radio Astronomy, Auf dem H\"ugel 69, 53121 Bonn, Germany
\and Universidade de Lisboa - Faculdade de Ci\^{e}ncias, Campo Grande, 1749-016 Lisboa, Portugal
\and Faculdade de Engenharia, Universidade do Porto, rua Dr. Roberto Frias, 4200-465 Porto, Portugal
\and CENTRA - Centro de Astrof\'isica e Gravita\c{c}\~{a}o, IST, Universidade de Lisboa, 1049-001 Lisboa, Portugal
\and Max Planck Institute for Astronomy, K\"onigstuhl 17, 69117, Heidelberg, Germany
\and Instituto de Astrof\'isica de Canarias (IAC), E-38200 La Laguna, Tenerife, Spain
\and Department of Astrophysical \& Planetary Sciences, JILA, University of Colorado, Duane Physics Bldg., 2000 Colorado Ave, Boulder, CO 80309, USA
\and Research School of Astronomy and Astrophysics, Australian National University, Canberra, ACT 2611, Australia
\and Instituto de Astronom\'ia, Universidad Nacional Aut\'oma de M\'exico, Apdo. Postal 70264, Ciudad de M\'exixo, 04510, M\'exico
\and Department of Physics, Technical University Munich, James-Franck-Stra{\ss}e 1, 85748 Garching, Germany
}

   \date{Received xxx, 2020; accepted January 29, 2021}
 
  \abstract{
   Using VLTI/GRAVITY and SINFONI data, we investigate the sub-pc gas and dust structure around the nearby type 1 AGN hosted by NGC~3783. The K-band coverage of GRAVITY uniquely allows a simultaneous analysis of the size and kinematics of the broad line region (BLR), the size and structure of the near-IR continuum emitting hot dust, and the size of the coronal line region (CLR). We find the BLR probed through broad \brg{} emission is well described by a rotating, thick disk with a radial distribution of clouds peaking in the inner region. In our BLR model the physical mean radius of 16 light days is nearly twice the 10 day time lag that would be measured, which matches very well the 10 day time lag that has been measured by reverberation mapping. We measure a hot dust FWHM size of 0.74 mas (0.14 pc) and further reconstruct an image of the hot dust which reveals a faint (5\% of the total flux) offset cloud which we interpret as an accreting cloud heated by the central AGN. Finally, we directly measure the FWHM size of the nuclear CLR as traced by the \caviii{} and narrow Br $\gamma$ line. We find a FWHM size of 2.2 mas (0.4 pc), fully in line with the expectation of the CLR located between the BLR and narrow line region. Combining all of these measurements together with larger scale near-IR integral field unit and mid-IR interferometry data, we are able to comprehensively map the structure and dynamics of gas and dust from 0.01--100 pc.
   }


   \maketitle
%
%-------------------------------------------------------------------
% Citation aliases
\defcitealias{GC2018Natur}{GC18}
\defcitealias{Gravity-Collaboration:2020ac}{GC20b}
\defcitealias{Gravity-Collaboration:2020ab}{GC20a}

\section{Introduction}
Recent advancements in infrared interferometric observations has allowed significant progress in understanding the gas and dust structure and dynamics around active galactic nuclei (AGN). With the new capabilities of GRAVITY \citep{GC2017FL}, the second generation instrument at the Very Large Telescope Interferometer (VLTI), the broad line region has been resolved and modelled to provide a measurement of the supermassive black hole mass \citep{GC2018Natur,Gravity-Collaboration:2020ab}, the hot dust has been imaged revealing a thin ring at the dust sublimation radius \citep{Gravity-Collaboration:2020aa}, and hot dust sizes have been measured for an increasing number of AGN \citep{Gravity-Collaboration:2020ac}. NGC~3783 provides a unique laboratory to study not only these aspects but also the coronal line region (CLR) in a single object and build a comprehensive picture of the nuclear and circumnuclear region around an AGN.

NGC\,3783 hosts one of the most luminous local AGN, with a bolometric AGN luminosity, $\log{L_{\rm AGN}} \sim 44.5$ erg s$^{-1}$ for a distance of 38.5\,Mpc \citep{Tully:1988aa,Davies:2015uq}\footnote{We adopt a distance of 38.5 Mpc based on the Tully-Fisher (TF) relation and reported on the NASA Extragalactic Database (NED) which is significantly smaller than the luminosity or angular size distance inferred from the redshift (47--48 Mpc) and due to the peculiar velocity of the galaxy. This distance though could potentially be affected by AGN contamination and thus underestimated. However, \citet{Crook:2007aa} found NGC~3783 to reside in a group of four galaxies at a distance of 37 Mpc and \citet{Kourkchi:2017aa} found NGC~3783 to reside in a group of nine galaxies at a distance of 42 Mpc. Therefore, we choose to use the TF based distance of 38.5 Mpc throughout our analysis and note the exact value of the distance within a few Mpc does not change the results.}, and is the subject of an extensive literature, most notably related to ionised outflows (especially X-ray warm absorbers) and variability.
As one of the most luminous AGN, NGC~3783 has been intensely monitored via reverberation mapping and is one of the few AGN 
for which simultaneous UV \citep{Reichert:1994aa} and optical \citep{Stirpe:1994aa} reverberation results have been obtained. NGC~3783 is also one of the AGN that clearly demonstrated the virial relationship between emission line lag and width \citep{Onken:2002aa}. Through reverberation mapping, the black hole mass of NGC\,3783 has been estimated to be $\sim3\times10^7$\,M$_\odot$ assuming a virial factor $f_\sigma =$ 4--5 \citep{Peterson:2004aa,Bentz:2015aa}, which implies that the AGN is radiating at $\sim0.1$\,L$_{\rm Edd}$.
For a velocity dispersion of 130\,km\,s$^{-1}$ as measured from both the Ca\,II triplet and the CO\,2-0 bandhead \citep{Caglar:2020aa}, this puts the object very close to the $M_{\rm BH}-\sigma*$ relation \citep{Ferrarese:2000gf,Gebhardt:2000xy,Ferrarese:2001aa,Nelson:2004aa,Onken:2004aa,Gultekin:2009ul,McConnell:2013fk}.
The black hole mass also matches that derived, together with the spin, from the Fe K$\alpha$ line at 6.4\,keV \citep{brenneman11,capellupo17}.

The X-ray warm absorber was initially modelled by \cite{netzer03} as requiring three different ionization components at two different velocities, at distances within limits of 0.2--25\,pc, and with a total column of N$_{\rm H} \sim 4\times10^{22}$\,cm$^{-2}$. 
With more detailed spectra, this was later expanded to three velocities  in the range $-460$ to $-1600$\,km\,s$^{-1}$ by \cite{mao19}, comparable to those identified previously in UV spectra of $-550$, $-720$, and $-1370$\,km\,s$^{-1}$ \citep{kraemer01}.
An X-ray obscuration event lasting about a month, which led to a significantly reduced flux at energies $\la5$\,keV and a modified ionisation structure, was reported by \cite{mehdipour17}.
Their modelling suggested this component had a column of N$_{\rm H}\sim 10^{23}$\,cm$^{-2}$, a covering factor of $\sim0.5$, and simultaneous UV observations \citep{kriss19} indicated it was moving out at a velocity of $\sim2000$\,km\,s$^{-1}$.
In addition, from the implied density of $3\times10^9$\,cm$^{-3}$ and similarity of the radial location at only 10\,light-days to the size of the broad line region (BLR) from reverberation mapping \citep{Onken:2002aa}, these authors argued that it is an obscuring wind originating in the outer part of the BLR.
A subsequent analysis of archival data by \cite{kaastra18} suggested that such obscuring events, with columns exceeding $5\times10^{21}$\,cm$^{-2}$ may be rather common for NGC\,3783.

Ionized outflows have also been observed on larger scales of tens to hundreds of parsecs \citep{Rodriguez-Ardila:2006aa}. 
Guided by the almost unresolved (<200\,pc diameter) appearance of the optical [OIII] image, \cite{Fischer:2013aa} modelled the kinematics in a slit spectrum as being consistent with an ionisation cone that is oriented 15$\degree$ from face-on with inner/outer opening angles of 45/55$\degree$ and a maximum outflow velocity of 130\,km\,s$^{-1}$.
In contrast, based on the biconical appearance of the near-infrared [SiVI] distribution at 16\,pc resolution, \cite{Muller-Sanchez:2011aa} modelled the line kinematics as an ionization cone that is $60\degree$ from face-on with inner/outer opening angles 27/34$\degree$ and a maximum outflow velocity of 400\,km\,s$^{-1}$. 
Neither of these are completely satisfactory: the former is more consistent with what is expected for a Seyfert~1 while the latter matches the expected outflow velocity better and explains the [SiVI] morphology.
However, face-on orientations appear to be ruled out by mid-infrared interferometric data which show strong elongation attributed to polar dust at a position angle of $-50$ to $-60\degree$ \citep{Honig:2013qy,Burtscher:2013aa,lopezgonzaga16}.
Fitting the near- to mid-infrared spectral energy distribution using a disk+wind model, \cite{hoenig17} reached a conclusion that is plausibly consistent with both, favouring an inclination of 30$\degree$ and an opening angle of 38$\degree$.
This would imply that in NGC\,3783 our line of sight is close to the edge of the ionisation cone, which could perhaps explain the slight extinction to the BLR (A$_{\rm V}\sim0.1$\,mag, \citealt{Schnorr-Muller:2016aa}), the classification as a Sy 1.2-1.5, and the frequent X-ray obscuration events.

These results indicate a polar axis that is oriented several tens of degrees west of north, consistent with that deduced from optical polarisation measurements.
Applying an equatorial disk model, \cite{smith02,smith04} concluded from polarisation data that the polar axis is at $-45\degree$. 
However, there are several issues that make this conclusion uncertain: NGC\,3783 does not show the expected position angle rotation across the H$\alpha$ line profile, the depolarisation in the line core may partially be due to the narrow lines, and the polarisation peak in the line wing is not unique to this specific model.
\cite{lira20} also noted that NGC\,3783 has unusual polarisation characteristics, with the position angle showing a more {\it M}-like profile possibly indicative of a radially outflowing scatterer. Further, both the mid-IR and polarisation measurements could be biased towards the edges of the outflow cones if this is the location of most of the mass in the cones and has been hinted at recently through observations of Circinus \citep[e.g.][]{Stalevski:2017aa,Stalevski:2019aa}.
Unfortunately, radio maps do not help in determining the orientation of the polar axis since they are unresolved at scales of 0.2--0.6\arcsec at 8.5\,GHz as well as 10--30\,mas at 1.6\,GHz \citep{Schmitt:2001aa,Orienti:2010aa}. Ironically, in both cases the beam is elongated in roughly the same direction as the polar axis, making it harder to assess whether there is a small scale radio jet.

To compound the issue surrounding the inner geometry of NGC\,3783, the orientation of the host galaxy disk is such that the kinematic major axis, measured from both the stellar and H$_2$ 1-0\,S(1) velocity fields on scales 30--300\,pc, are consistent with $-30$ to $-40\degree$ \citep{Davies:2007aa,Hicks:2009aa,Lin:2018aa} which matches the kinematic axis on kiloparsec scales measured from H$\alpha$ in a recent VLT/MUSE observation \citep{den-Brok:2020aa}.
These are very similar to the $-45\degree$ kinematic major axis on tens of kiloparsecs measured from HI \citep{garciabarreto99}, suggesting that over most of the galaxy disk there is little warping.
It means that the host galaxy kinematic axis and the outflow direction, at least on scales of a few tens of parsecs and more, are both oriented north to north-west, and cannot provide a guide for disentangling the innermost geometry of this galaxy.

In this paper we present new data from the VLTI spectrometer GRAVITY \citep{GC2017FL} which combines the light from all 4 UTs of the VLT, and can resolve structures on scales smaller than a few milliarcsec and  provide spectro-astrometric relative precision of tens of microarsec.
This resolution is essential to resolve the broad line region and the hot dust distribution immediately around it.
Analysis of the H- and K-band variability by \cite{lira11} revealed a 70\,day lag with respect to optical continuum, indicative of a radius of 0.06\,pc for the hot dust distribution. 
The size measured from K-band interferometry is very comparable at 0.1\,pc \citep{Weigelt:2012aa,Gravity-Collaboration:2020ac}.
The new data we report here present a new look at this innermost region of NGC\,3783.

This work adopts the following parameters for a $\Lambda$CDM cosmology: 
$\Omega_m = 0.308$, $\Omega_\Lambda = 0.692$, 
and $H_{0}=67.8$ km s$^{-1}$ Mpc$^{-1}$ \citep{Planck2016AA}. Using this 
cosmology and our adopted distance of 38.5 Mpc, 1~pc subtends 5.36~mas on sky and 1~$\mu$as corresponds to 0.22 light days.

%--------------------------------------------------------------------
\section{Observations and Data Reduction}
\subsection{GRAVITY}
We observed NGC~3783 with GRAVITY \citep{GC2017FL} over 6 nights spanning 3 years through a series of Open Time programmes and our Large Programme\footnote{Observations were made using the ESO Telescopes at the 
La Silla Paranal Observatory, program IDs 0100.B-0582, 0101.B-0255, 0102.B-0667, 2102.B-5053, and 1103.B-0626.} aimed at measuring the size of the BLR and the mass of the supermassive black hole. We used single-field on-axis mode with combined polarisation and medium ($R\sim500$) spectral resolution in the science channel for all observations. In single-field on-axis mode, both the fringe tracker (FT) and science channel (SC) fibres are centred on the same object with each receiving 50\% of the light. A single exposure lasted 5 min and consists of 1) coherent 3.3ms integrations with the FT fibre and 2) a series of 10 exposures with detector on-chip integration times (DITs) of 30s with the SC fibre. Exposures were done in sequence with only intermittent sky and calibrator star observations.  Table~\ref{tab:obs} provides a log of the GRAVITY observations used for our analyses in this paper including the range of seeing and coherence time throughout the observations. 

%************************************
% Log table for Gravity observations
\input{tab1}
%************************************

We reduced all data using the latest version of the GRAVITY pipeline \citep{Lapeyrere2014SPIE}. For the FT continuum visibility data we applied the default settings. For the SC data, we followed \citet[][hereafter GC20a]{Gravity-Collaboration:2020ab} and \citet[][hereafter GC20b]{Gravity-Collaboration:2020ac} and chose to retain all SC DITs regardless of FT SNR (\texttt{snr-min-ft = 0}) and estimated visibility loss (\texttt{vfactor-min-sc = 0}) which we found results in a substantial improvement in the phase noise. 

After the pipeline reduction of the FT data, we further applied the selection method first presented in \citetalias{Gravity-Collaboration:2020ac} where only FT DITs with a group delay $<3\mu{\rm m}$ are selected before averaging together and then calibrating using the calibrator star observations. This improves the averaged visibility squared data and removes its dependence on the Strehl ratio. 

For the SC data, we applied the additional steps outlined in \citetalias{Gravity-Collaboration:2020ab} to improve the SC differential phase spectra. This involves first fitting and subtracting an instrumental phase model that accounts for residual phase features introduced as a result of the 3rd order polynomial used in the pipeline being insufficient to flatten the phase spectra below the 1\degree{} level. As a final step, we fit and subtract a local first order polynomial around the \brg{} line to further refine the flattening. SC differential amplitude spectra are produced by fitting and dividing out a 2nd order polynomial from the pipeline reduced visibility amplitude spectra. We estimate the uncertainty for the differential phase and amplitude spectra by calculating the RMS in the line-free regions.

\begin{figure}
\centering
\includegraphics[width=\columnwidth]{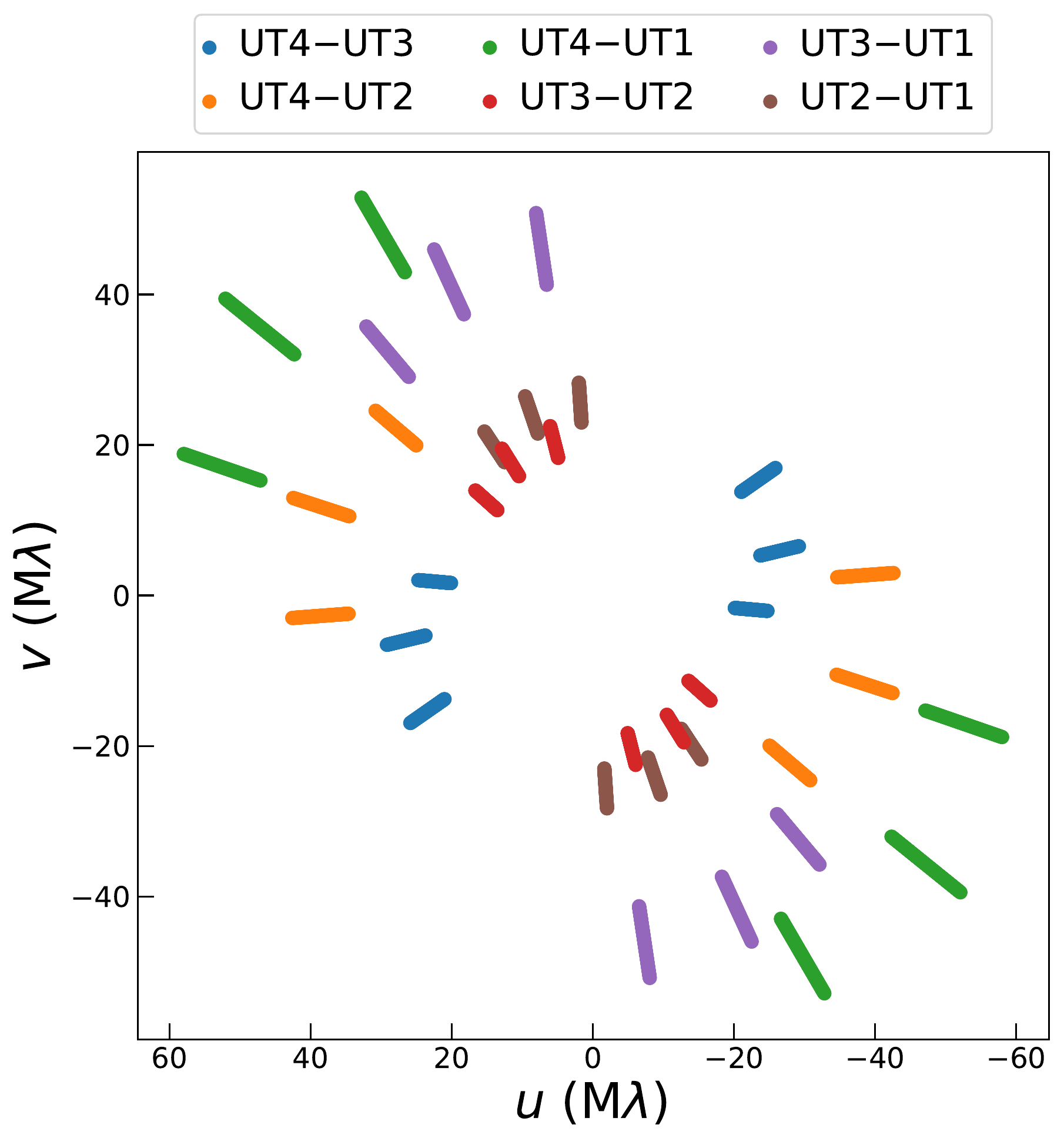}
\caption{$uv$ coverage of the binned GRAVITY observations. The radial span of each stripe is due to the wavelength range of the data. $uv$ bins 1, 2, and 3 are ordered in clockwise direction for each baseline.}
\label{fig:uvbin_cov}
\end{figure}

To still further improve the SNR of the SC data, we chose to bin and stack the data based on their location in the $uv$ plane. We split the data into three angular $uv$ bins such that each bin covers one third of the full $uv$ track for each baseline. We stack all SC data within a $uv$ bin using the inverse of the squared RMS uncertainty as the weight. Fig.~\ref{fig:uvbin_cov} shows the average $uv$ coordinates for each bin with extensions in the radial direction due to the spectral coverage of GRAVITY. 
Appendix~\ref{app:uv_bin_spec} shows the full $uv$-binned spectra, however the differential phase spectra have had the ``continuum phase'' signal subtracted (see Sec.~\ref{sec:blr_model}).

\subsection{SINFONI}

We observed NGC~3783 on April 20, 2019 using SINFONI with adaptive optics in service mode at the VLT \citep{Eisenhauer:2003eh, Bonnet:2004aa}. The observations were obtained at the $25$ mas pix$^{-1}$ scale and the $K$ grating ($R \simeq 4000$). A total of 6 exposures of $100$s were taken in a dithering object-sky-object sequence. Data were reduced with our standard pipeline and wavelength calibration scheme \citep[SPRED,][]{Schreiber:2004aa,Abuter:2006aa}. Sky frames were subtracted from the images to correct for instrumental and atmospheric background. We then reconstruct $12.5$ mas pix$^{-1}$ images with a spectrum at each pixel, applying flat-fielding, bad-pixel, distortion, and cosmic-ray hit corrections. The wavelength calibration is done using emission line lamps and is further tuned on the atmospheric OH lines in the raw frames. The individual data cubes were then combined, and a telluric correction was applied using \texttt{molecfit} \citep{Smette:2015aa,Kausch:2015aa}. Finally, to improve the S/N in the outer regions of the field of view (FOV) to derive emission line maps, in particular of the \caviii{} line, we smoothed the data cube with a 6 pixel FWHM Gaussian in the spatial directions, equal to the 75 mas FWHM point spread function (PSF) measured from a fit to the continuum\footnote{We further checked the PSF FWHM using our fit to the broad component of \brg{} which also is unresolved and found similar FWHM both before and after smoothing.}, and a 2 pixel FWHM Gaussian in the spectral direction, matching the instrumental dispersion.

\begin{figure*}
    \centering
    \includegraphics[width=0.8\textwidth]{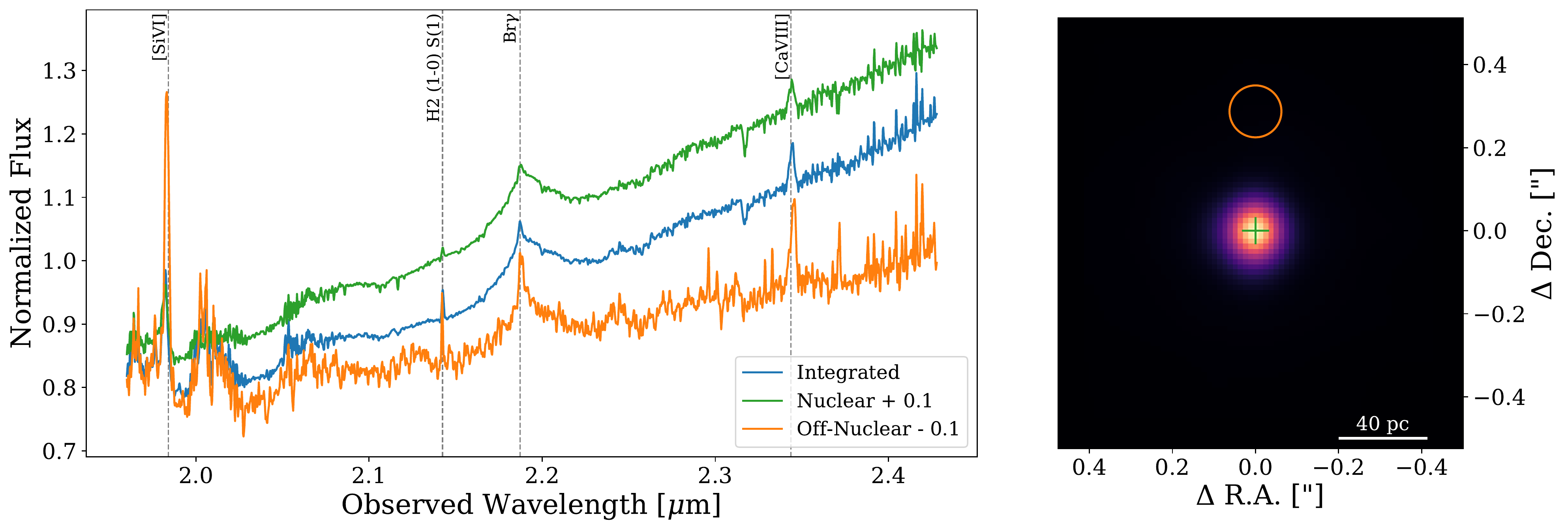}
    \caption{\textit{Left:} Example spectra extracted from the SINFONI cube including the full integrated spectrum (blue), a spectrum from a nuclear spaxel (green), and an integrated spectrum from an off-nuclear region (orange). All spectra have been normalized by their median flux and slightly offset to improve visualisation. \textit{Right:} K-band continuum image created by integrating the SINFONI cube between 2.25 and 2.31 \micron. The green cross indicates the spaxel used for the nuclear spectrum in the right panel. The orange circle indicates the aperture used for the off-nuclear spectrum.}
    \label{fig:sinfo_spec}
\end{figure*}

Fig.~\ref{fig:sinfo_spec} plots example spectra (left panel) from the smoothed SINFONI cube including (1) the full integrated spectrum, (2) a spectrum from a nuclear spaxel at the position of the green cross in the continuum image (right panel), and (3) a spectrum from an off-nuclear region indicated by the orange circle in the continuum image. We further show the expected locations of the \sivi, H$_{2}$ (1--0) S(1), \brg, and \caviii{} emission lines.

\subsection{The normalised \brg{} and \caviii{} profiles}
\begin{figure}
    \centering
    \includegraphics[width=\columnwidth]{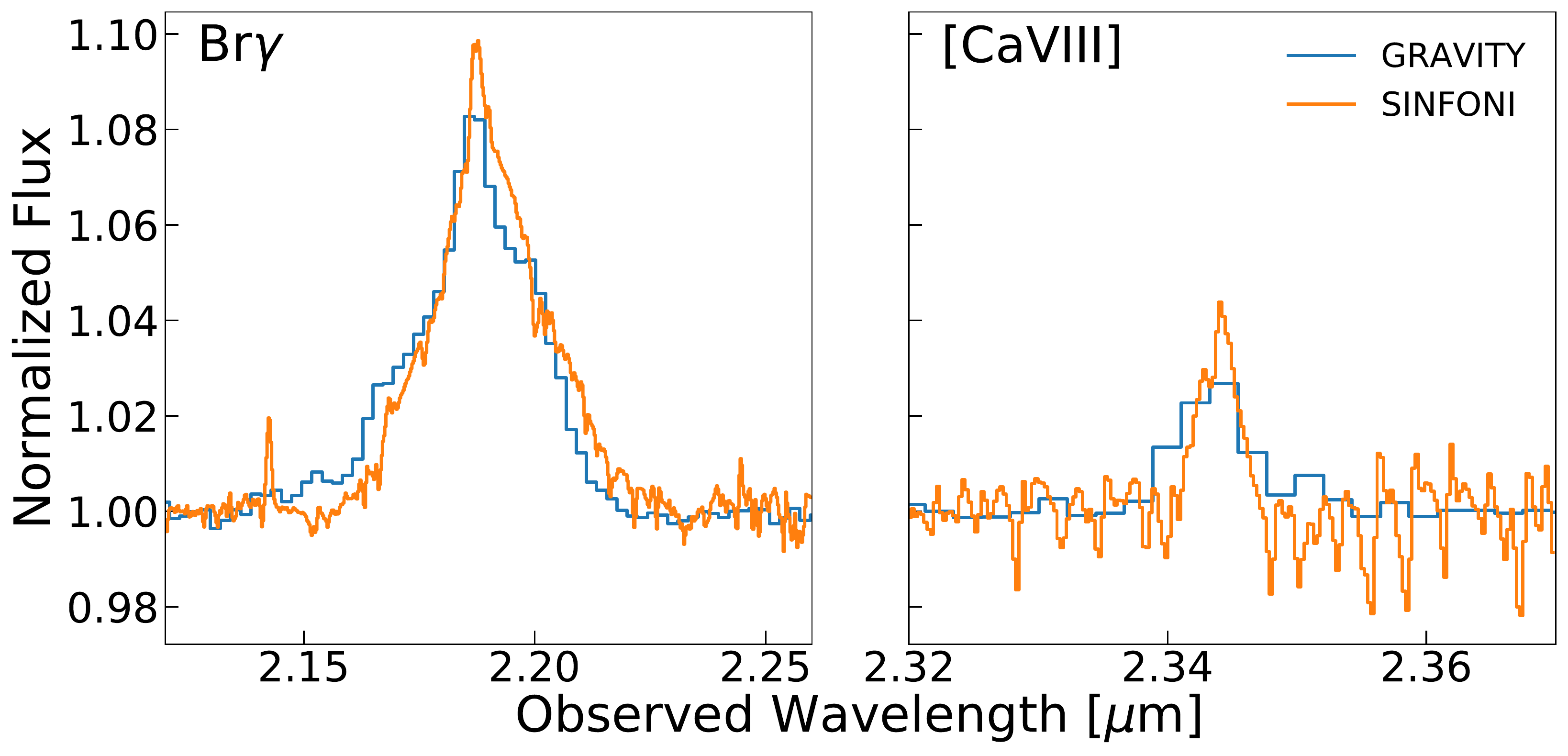}
    \caption{The normalised \brg{} and \caviii{} line profiles observed by GRAVITY (blue) and SINFONI (orange).}
    \label{fig:flux_spec}
\end{figure}

Besides the interferometric observables (e.g. differential phase and amplitude, closure phases, etc), an important component of our analysis is the emission line flux profile normalised to the continuum which is also measured by GRAVITY. Following \citetalias{Gravity-Collaboration:2020ab}, to construct the normalised line profiles, we only used observations where an early type star (Feb. 16, 2019 and Mar. 08, 2020) was observed as a calibrator to correct for telluric features. Using an early type star avoids the complicated stellar absorption features that are prominent in late type stars above 2.3 \micron{} where we expect the faint \caviii{} line. However, early type stars have strong \brg{} absorption which occurs in the blue wing of NGC~3783's broad \brg{} profile. Fortunately, these calibrators were also observed on the same night as IRAS 09149-6206, whose \brg{} emission does not overlap with the calibrator's \brg{} absorption. Thus, we fit a line profile to the calibrator's absorption in the IRAS 09149-6206 spectrum from each night and used these fits to remove the stellar \brg{} feature from the NGC~3783 spectra. The spectra from each of the two nights were then averaged together, weighted by their uncertainties to produce the final GRAVITY flux spectrum. Fig.~\ref{fig:flux_spec} shows the final GRAVITY mean spectrum for both \brg{} and \caviii{}.

%***************SINFONI Spectra Figure*******************************************************
%*****************************************************************************************

%***************Flux Spectrum Figure*******************************************************
%*****************************************************************************************

With higher spectral resolution and S/N, the SINFONI data also provide a high quality K-band spectrum that can be compared with the GRAVITY spectrum. We extracted a nuclear spectrum using a circular aperture centred on the peak of the continuum and a diameter of 9 pixels that matches the PSF FWHM of the smoothed cube. The region around the \brg{} and \caviii{} lines were then normalised to a local fit of the continuum. Fig.~\ref{fig:flux_spec} plots the normalised \brg{} and \caviii{} profiles along with the GRAVITY profiles.

For both \brg{} and \caviii{}, the SINFONI spectra show slightly higher peak normalised fluxes. This can be explained by the SINFONI spectrum covering a much larger physical area compared to GRAVITY. Whereas the FOV of GRAVITY is $\sim60$ mas, the SINFONI spectrum is produced by integrating over a circle with a diameter of 112.5 mas. Narrow \brg{} emission, which is largely contributing to the peak of the \brg{} profile, and \caviii{} emission is expected to occur over a large range of size scales. The additional area covered by SINFONI then increases the relative flux of the narrow \brg{} and \caviii{} lines since the continuum is much more compact.

Our choice of which line profile to use depends on the specific analysis we want to perform. For our BLR modelling (Sec.~\ref{sec:blr_model}), we choose to use the SINFONI profile because of its much higher spectral resolution. Small deviations from a smooth line profile can be important and signal BLR substructure. For our coronal line region (CLR) analysis, we use the GRAVITY profiles because they represent the correct line-to-continuum ratio that matches the interferometric data. 

\begin{figure}
    \centering
    \includegraphics[width=0.8\columnwidth]{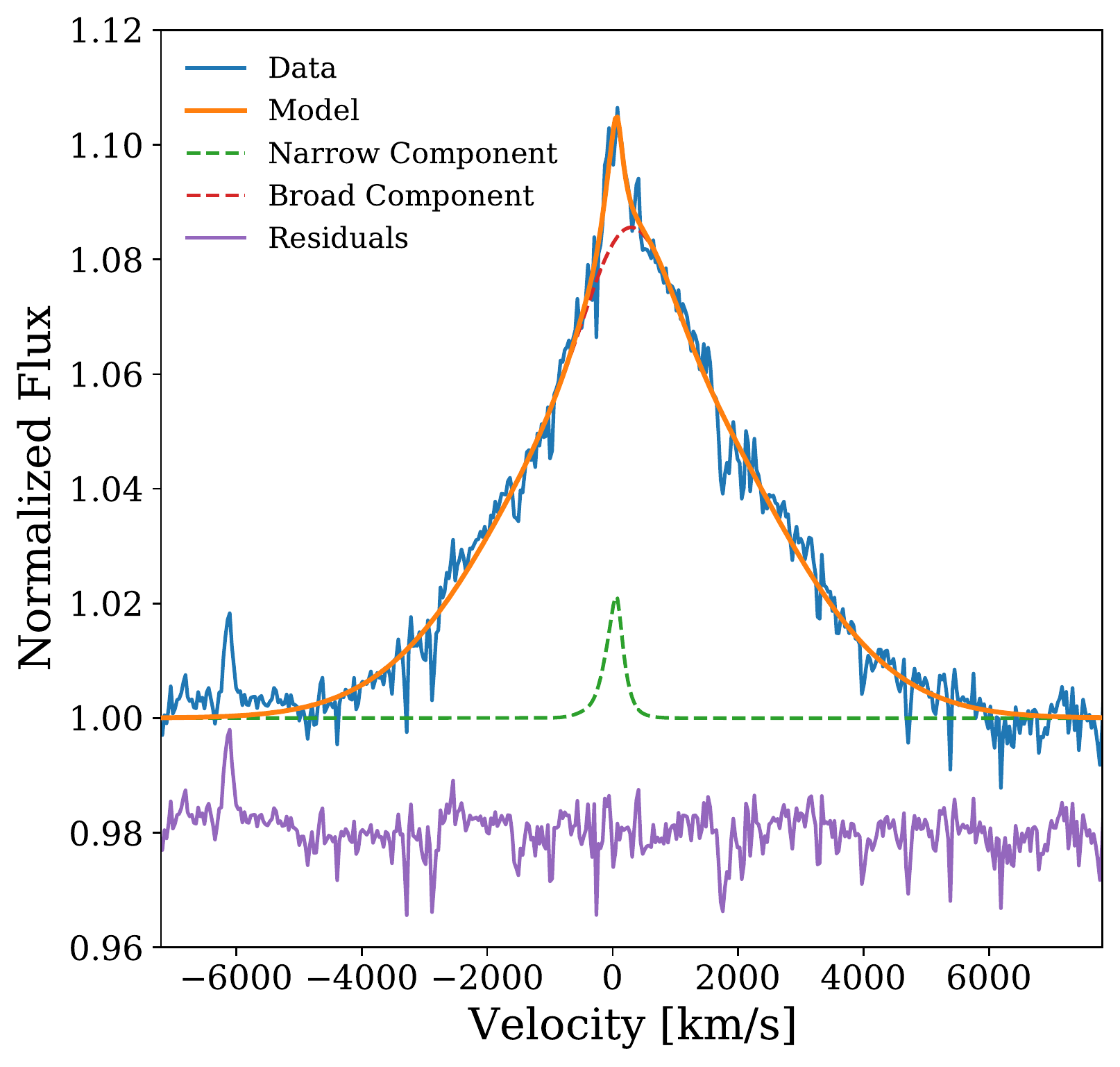}
    \caption{Normalized \brg{} profile from SINFONI (blue) with the best fit line profile model (orange). The model is the combination of a narrow line template determined by the [\ion{O}{iii}]$\lambda5007$\AA{} line from X-Shooter spectra (green) and a broad component consisting of 2 Gaussians (red). The residuals (purple) show that our fit very well matches the observed \brg{} profile.}
    \label{fig:sinfoni_line_fit}
\end{figure}

For the BLR analysis, we further have to account for the narrow \brg{} component. To be as accurate as possible, we use the [\ion{O}{iii}]$\lambda5007$\AA{} line profile as a template of the narrow line. To produce the template, we fit the [\ion{O}{iii}] line from previous X-Shooter spectra \citep{Schnorr-Muller:2016aa} with 4 Gaussian components that were needed to accurately model the line. The velocities, widths, and relative amplitudes of these Gaussian components were then fixed to create the narrow line template. The template was broadened to account for the spectral resolution difference between X-Shooter and SINFONI. This broadened template along with 2 Gaussian components to describe the broad component were fit to the SINFONI \brg{} spectrum with only an overall scale factor and velocity shift for the narrow line template allowed to vary. Fig.~\ref{fig:sinfoni_line_fit} plots the results of our line fitting and shows that this model very well reproduces \brg{} profile. The narrow component was then subtracted from the spectrum to produce the profile we will use for our BLR analysis. 

\section{The structure of the BLR and the SMBH mass}

\begin{figure*}
\centering
\includegraphics[width=0.7\textwidth]{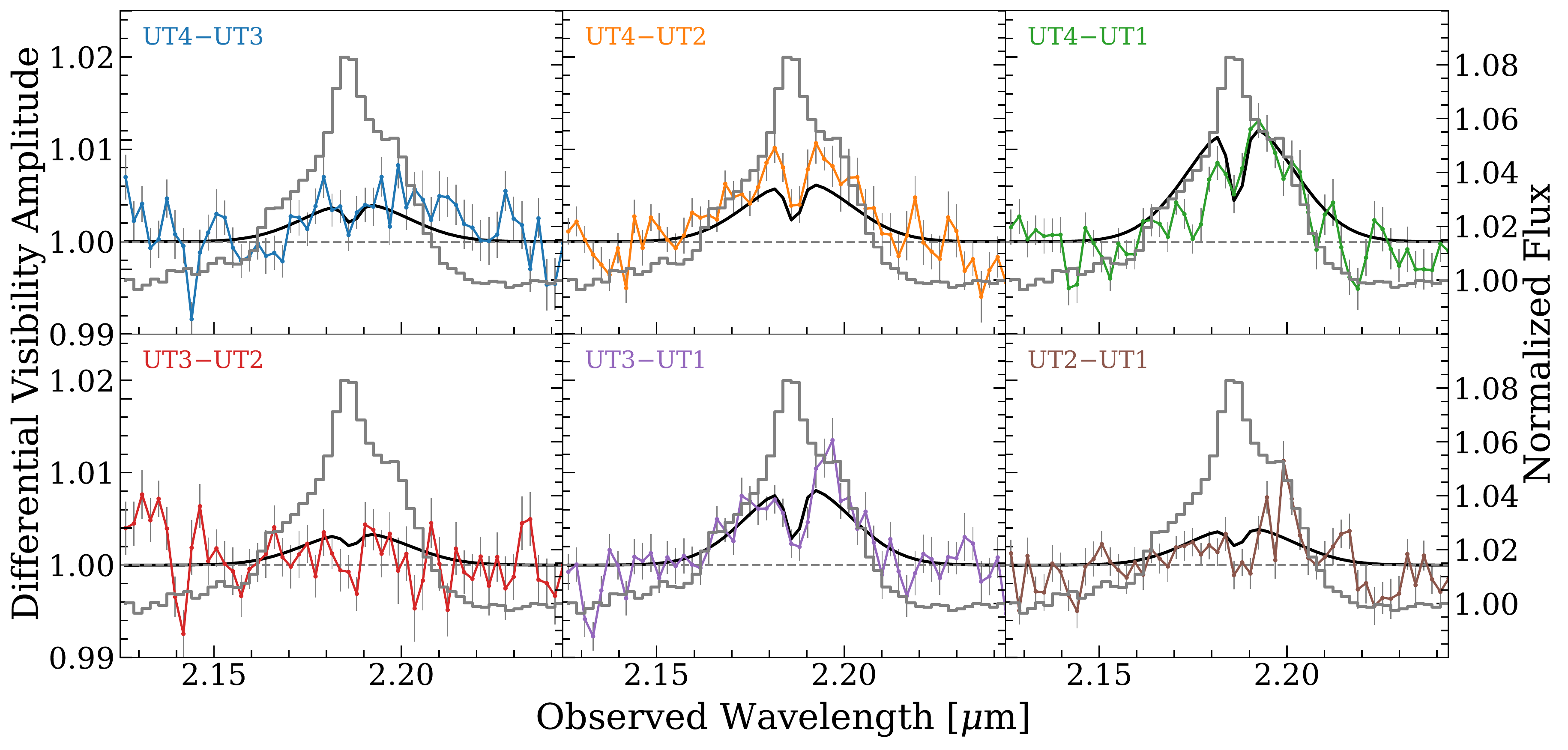}
\caption{Baseline averaged differential visibility amplitude spectra spanning the \brg{} line (coloured points and grey error bars). The grey profile in each panel plots the same normalised \brg{} spectrum from GRAVITY. The black curves represent the best fit model from our CLR analysis in Sec.~\ref{sec:clr}.}
\label{fig:sc_visamp}
\end{figure*}

\begin{figure*}
\centering
\includegraphics[width=0.7\textwidth]{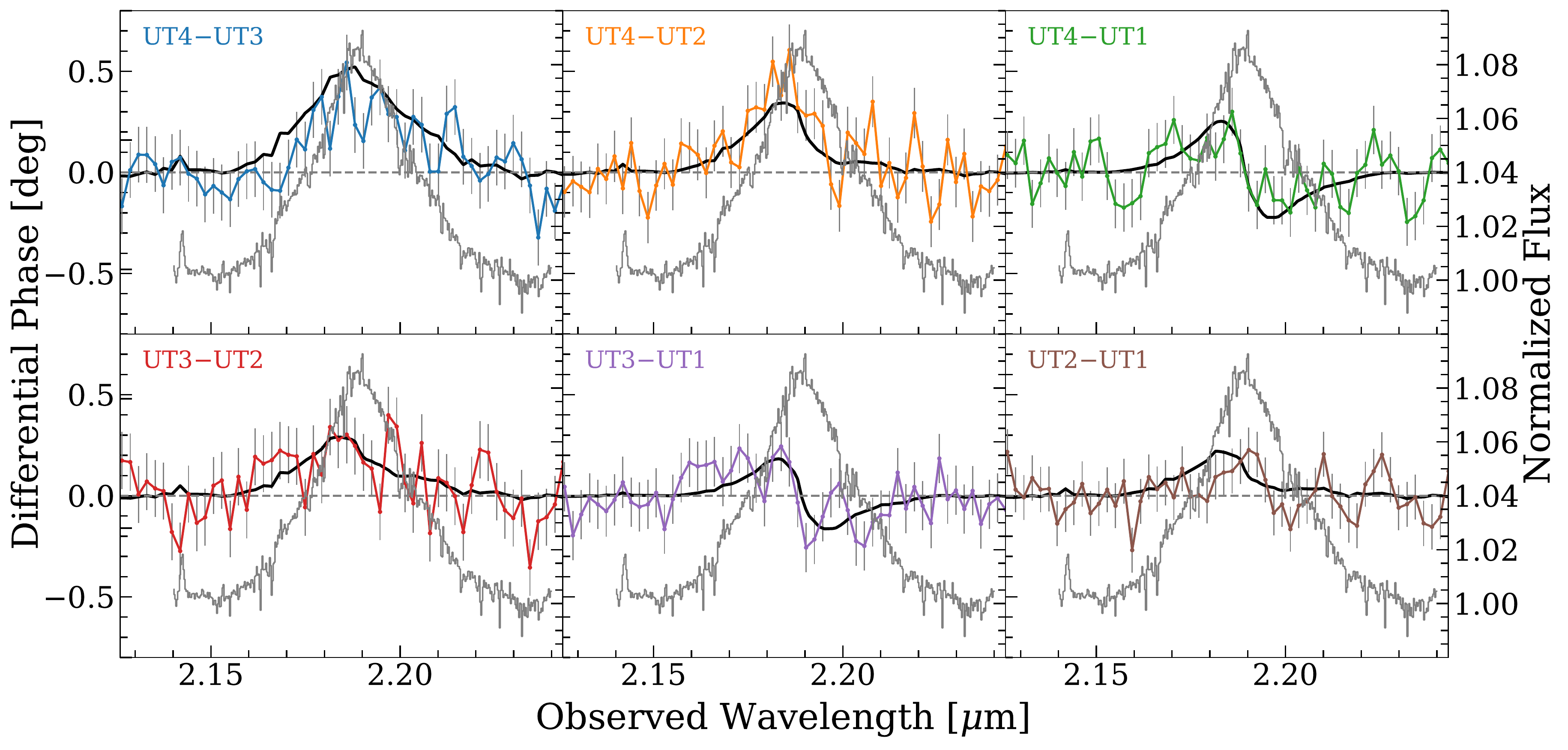}
\caption{Baseline averaged differential phase spectra spanning the \brg{} line (coloured points and grey error bars). The grey profile in each panel plots the same normalised \brg{} spectrum from SINFONI with the narrow component removed. The black curves represent the best fit BLR model summed with the continuum phase calculated from our image reconstruction. The model line profile looks noisy due to the continuum phase contribution being calculated based on the observed line profile.}
\label{fig:sc_visphi}
\end{figure*}

The baseline averaged differential visibility amplitude spectra plotted in Fig.~\ref{fig:sc_visamp} definitively show we have detected the BLR. The positive bumps ($\Delta V > 1$) seen in the longest baselines (UT4$-$UT2, UT4$-$UT1, and 
UT3$-$UT1) over the \brg{} line indicate a smaller BLR size compared to the continuum, as expected. While the differential phase spectra are noisier, we also observe positive bumps ($\Delta \phi > 0$) in the UT4$-$UT3 and UT4$-$UT2 baselines and an ``S-shape`` signal in the UT4$-$UT1 and UT3$-$UT1 baselines as shown in Fig.~\ref{fig:sc_visphi}. As seen for IRAS 09149-6206 (\citetalias{Gravity-Collaboration:2020ab}), a differential phase signal following the line profile (i.e. 
``continuum phase'') is produced from a difference in interferometric phase between the BLR and hot dust continuum. An ``S-shape'' signal, on the other hand, can be produced by ordered rotation such as that detected in 3C 273 \citep{GC2018Natur}. 

In contrast to IRAS 09149-6206 though, we are able to directly constrain the ``continuum phase'' contribution through our image reconstruction of the hot dust continuum (see Sec.~\ref{sec:image_reconstruct}). To do this, we first simulate differential phase GRAVITY data for each observation based on our best image reconstruction by running it through our in-house built GRAVITY simulator. This produces the complex phase signal associated with the hot dust continuum structure. We then calculate the differential phase signal by multiplying the complex phase by $-2\pi f_\lambda/(1 + f_\lambda)$ where $f_\lambda$ is the line flux at wavelength $\lambda$ relative to a continuum level of unity. These ``continuum phase`` spectra are finally subtracted from the original differential phase spectra. The subtracted differential phase spectra are shown in Fig.~\ref{fig:uvbin_bry_visphi}. In the following sections we use these spectra to measure photocentres and fit a dynamic BLR model.

\subsection{Photocentre Fitting}

%******************** Photocentre Figure **************************************************
\begin{figure*}
    \centering
    \includegraphics[width=\textwidth]{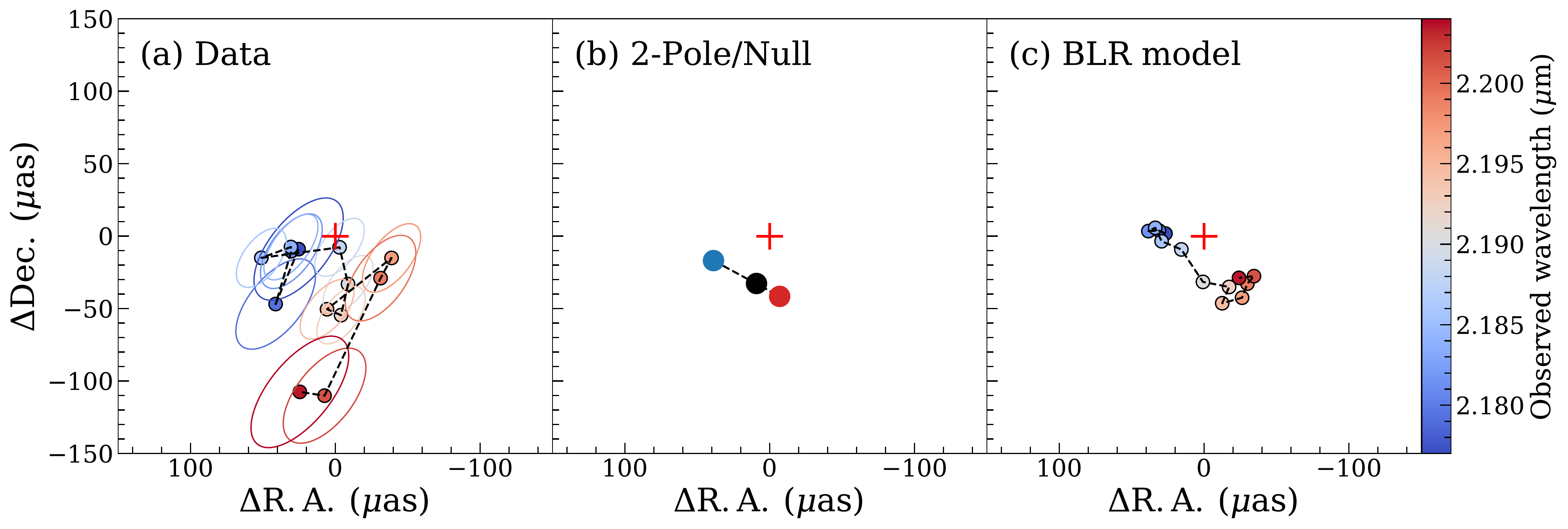}
    \caption{(a) Best fit photocentres for the 13 brightest spectral channels covering the \brg{} line to the differential phase data. The colour indicates wavelength and the ellipses outline the uncertainty on the photocentres. The red cross plots the (0,0) position. (b) Best fit ``2-pole'' photocentre model where all blueshifted (blue point) and redshifted (red point) channels are assumed to have the same photocentre with the addition of the wavelength independent offset. The black point indicates the best fit position of the ``null'' model where all channels are assumed to have the same photocentre. (c) Photocentres produced from our best fitting BLR model (see Section~\ref{sec:blr_model}) which we calculated by first producing mock differential phase spectra and fit them in the same way as panel (a). Colours are the same as in panel (a).}
    \label{fig:photocent}
\end{figure*}
%******************************************************************************************

A first test of our qualitative assessment of the BLR structure is calculating the photocentres of the spectral channels where \brg{} line emission dominates. Following our analysis of IRAS 09149-6206 (\citetalias{Gravity-Collaboration:2020ab}), we fit the continuum phase subtracted, $uv$-binned differential phase spectra with the following equation,

\begin{equation}\label{eq:diff_phase_photocent}
\Delta \phi_\lambda = -2\pi\, \frac{f_\lambda}{1 + f_\lambda}\, 
\vec{u} \cdot \vec{x}_\mathrm{BLR,\lambda},
\end{equation}

\noindent where $\vec{u}$ is the 
$uv$ coordinate of the baseline, $\vec{x}_{\mathrm{BLR},\lambda}$ is the 
BLR coordinate of the photocentre at wavelength $\lambda$ w.r.t. the photocentre of the continuum.

We fit spectral channels with $f_\lambda > 1.04$ resulting in 13 model-independent photocentres that are shown in Fig.~\ref{fig:photocent}a. Interestingly, even though the continuum phase has already been subtracted off, we still see a slight residual systematic offset of the photocentres from the origin. This is likely caused by our relatively large pixel scale (100 \uas{}) in our image reconstruction. Any slight shift of the continuum from the centre on scales less than 1 pixel would not be captured in the reconstructed continuum phase. Indeed, the residual offset is much less than 100 \uas{}.  
Despite the small systematic shift, the photocentres still show a general NW--SE velocity gradient which would cause the ``S-shape'' signal in other baselines. 

We further test the robustness of the velocity gradient by fitting for a single photocentre for all 5 blueshifted channels and all 8 redshifted channels relative to the line centre of 2.1866 \micron, our ``2-pole'' model. The blue and redshifted ``poles`` are shown in Fig.~\ref{fig:photocent}b as blue and red dots respectively. As in the individual photocentre fitting, we still find the general velocity gradient in the same direction. We use an F-test to estimate the significance of the gradient by comparing to a null hypothesis where all of the spectral channels are located at a single photocentre (i.e. 
``null'' model). The best fit photocentre of the null model is shown as a black dot in Fig.~\ref{fig:photocent}b. Comparing the 2-pole and null models, we find the detection of the velocity gradient is significant at $>8\sigma$. Using the separation of the blue and red poles, we can also place a rough estimate on the size of the BLR with $R_{\rm BLR} \sim 30$ \uas, consistent with the $\sim10$ light-day time lags measured from reverberation mapping \citep{Stirpe:1994aa,Onken:2002aa,Peterson:2004aa,Zu:2011aa} However, to truly measure the size of the BLR and its other properties we need to fit the data with a physical model for the BLR. 

\subsection{BLR modelling}\label{sec:blr_model}

\begin{figure*}

\begin{minipage}{.5\textwidth}
\centering
\subfloat[]{\includegraphics[height=0.3\textheight]{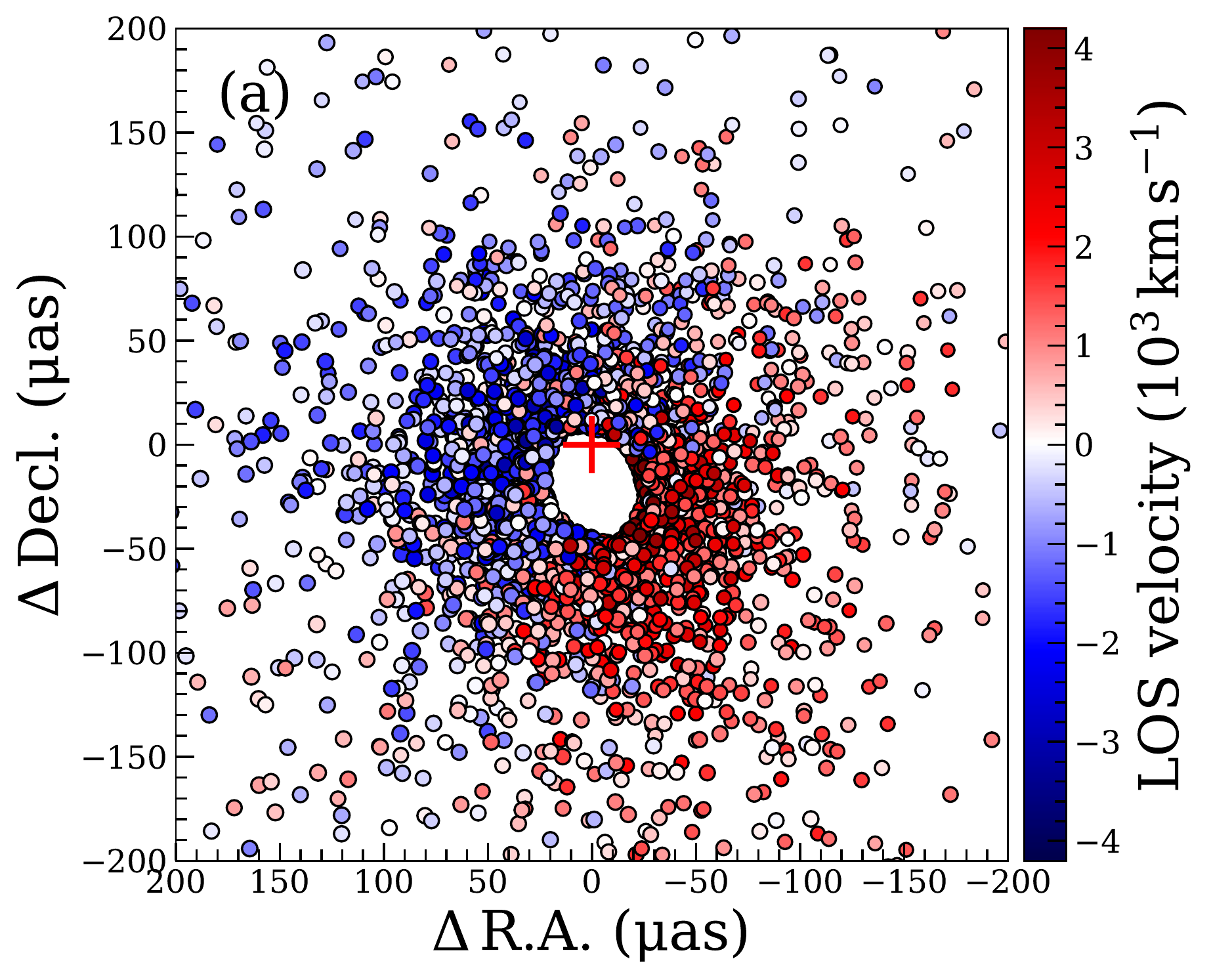}}
\end{minipage}%
\begin{minipage}{.5\textwidth}
\centering
\subfloat[]{\includegraphics[height=0.3\textheight]{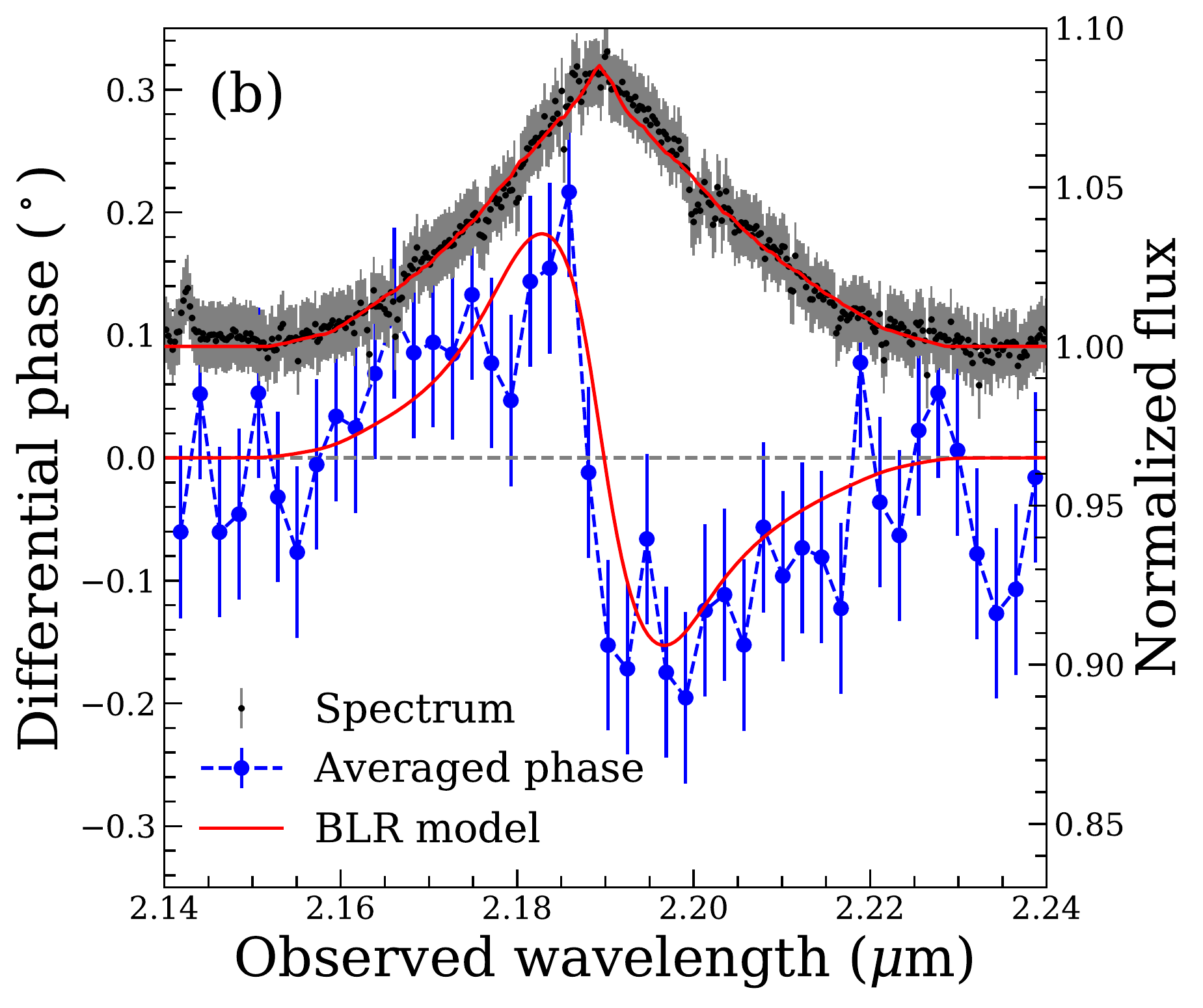}}
\end{minipage}\par\medskip
\centering
\subfloat[]{\includegraphics[height=0.3\textheight]{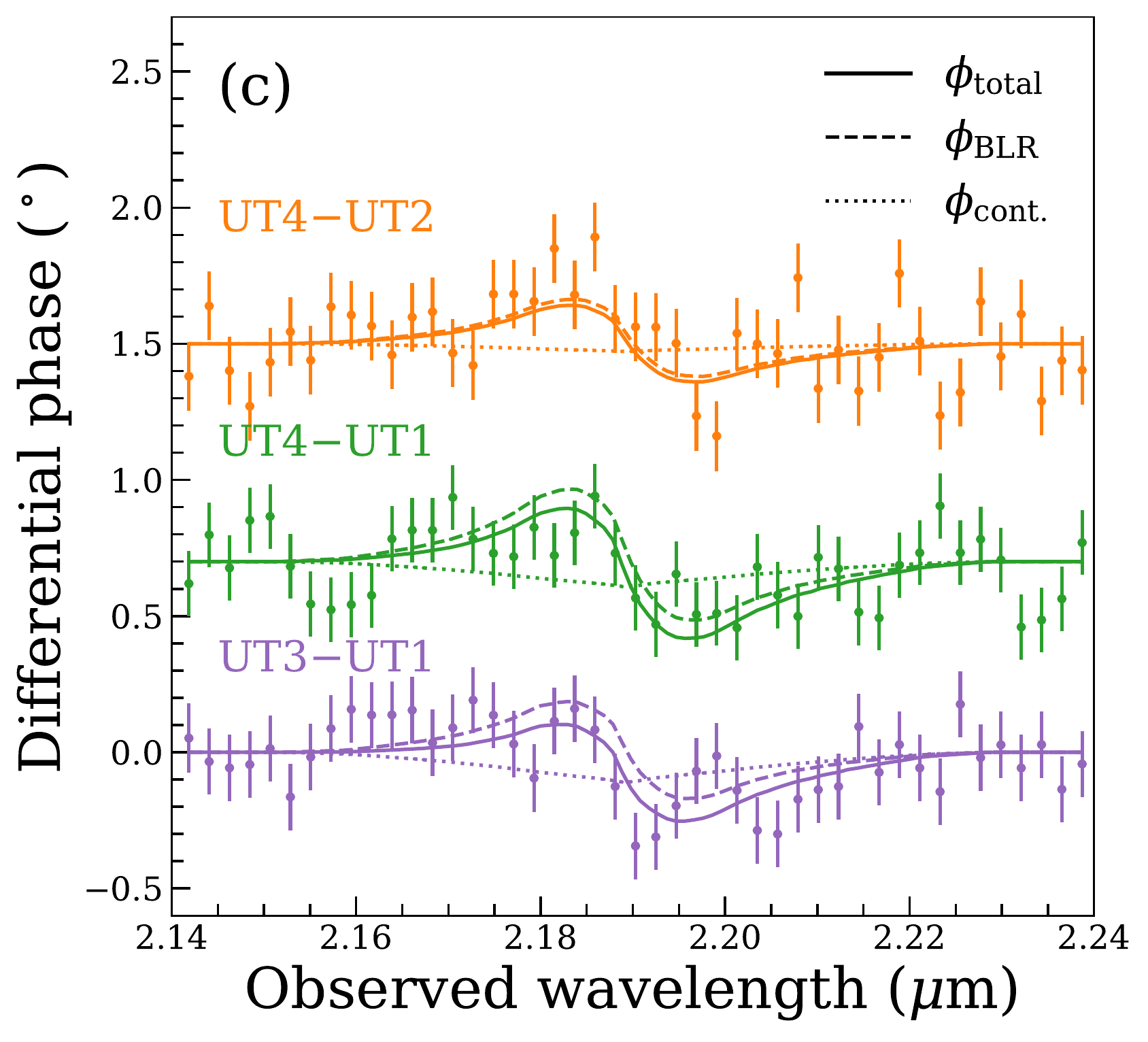}}

\caption{(a) The cloud distribution of the best-fit BLR 
model. Each circle represents one cloud, colour-coded by the line-of-sight 
velocity. The green ellipse at the origin illustrates the uncertainty of the offset of the 
BLR centre. 
(b) The observed averaged differential phase from baselines UT4$-$UT2, UT4$-$UT1, and 
UT3$-$UT1 after removing the residual `continuum phase' signal (blue points) compared to the 
averaged differential phase from the best-fit BLR model (lower red line). The black points and overlaid red line show the observed and best model line flux profile respectively.
(c) The averaged differential phase data (points) and the best-fit models (lines) of the 
three baselines that show the strongest signal of the BLR component (dashed 
lines).  The phase in panel (b) is calculated by averaging the phases of these 
three baselines after subtracting the best-fit residual continuum phases (dotted lines). }
    \label{fig:blr_model_circ}

\end{figure*}

%*********************************************************************************************

Given the high significance of the velocity gradient and to constrain the properties of the BLR, we choose to fit our data with a physical model of the BLR. We use the model first developed in \citet{Pancoast2014MNRASa} and updated for GRAVITY data in \citet{GC2018Natur} and \citetalias{Gravity-Collaboration:2020ab}. As \citetalias{Gravity-Collaboration:2020ab} contains a detailed description of the BLR model, we only provide here a brief description necessary for our analysis. The underlying assumption of the model is that the BLR is composed of a large number of non-interacting clouds under the gravitational influence of a central SMBH with mass, \mbh. The radial distribution of the clouds is described by a shifted gamma distribution with a hard lower limit of the Schwarzschild radius and free parameters; $\beta$, controlling the shape of the distribution; $F$, the fractional inner radius; and $R_{\mathrm{BLR}}$ the mean BLR radius. Clouds are then distributed randomly both around the rotation axis and above the midplane up to the angular thickness of the disk, $\theta_0$. 

Three different parameters ($\kappa$, $\gamma$, $\xi$) introduce asymmetry in the BLR emission: $\kappa$ controls the fraction of line emission emitted from individual clouds towards or away from the central source and hence the fractional emission in the observer's direction from the near and far sides of the BLR; $\gamma$ controls whether the outer surface of the BLR preferentially emits line emission; and $\xi$ controls the level of midplane transparency. Together these three parameters control the relative weight of each cloud in determining the total emission.

Each cloud is placed on a bound elliptical orbit to model the kinematics. The tangential ($v_{\phi}$) and radial ($v_r$) velocities are randomly chosen based on a distribution centred around $v_{\phi} = v_{\mathrm{circ}}$ and $v_r = 0$ with $v_{\mathrm{circ}}$ equal to the Keplerian circular velocity given the cloud's radial position and \mbh. The distribution follows an ellipse described by Equation 6 of \citetalias{Gravity-Collaboration:2020ab} and has a Gaussian shape with standard deviations, $\sigma_{\mathrm{\Theta, circ}}$ along the ellipse and $\sigma_{\mathrm{\rho, circ}}$ perpendicular to the ellipse. A fraction of the clouds ($1 - f_{\rm ellip}$) can be placed on highly elongated orbits dominated by radial motion. Whether these clouds are inflowing or outflowing is controlled by $f_{\rm flow}$ where $f_{\rm flow} < 0.5$ indicates inflow and $f_{\rm flow} > 0.5$ indicates outflow.

The cloud distribution is then rotated on sky by an inclination angle, $i$, and position angle, $PA$ and translated by an offset, $(x_0, y_0)$. Line-of-sight (LOS) velocities are calculated including both the full relativistic Doppler effect and gravitational redshift. Finally, clouds are binned into spectral channels according to their LOS velocity. The flux for each spectral channel is the sum of the weights for all clouds within the bin and the photocentre is the weighted average position on sky of all the clouds within the bin. The model flux profile is normalised such that the maximum is $f_{\mathrm{peak}}$ and differential phases are then calculated according to Equation~\ref{eq:diff_phase_photocent}. Both the model flux profile and differential phase spectra are compared to the observed flux profile and differential phase spectra and the model parameter posterior distributions are sampled using nested sampling with the \texttt{dynesty} Python package \citep{Speagle2020MNRAS}. Priors on each parameter are the same as given in \citetalias{Gravity-Collaboration:2020ab}.

%************************************
% Table for best fit model parameters
\input{tab2}
%************************************

Table~\ref{tab:blr} lists the best fit parameters and uncertainties from our BLR modelling and Fig.~\ref{fig:corner_blr} shows the posterior distributions. Best fit parameter values are determined from the maximum \textit{a posteriori} values of the joint posterior distribution. Uncertainties on these values represent the 95\% credible interval estimated from the marginalised posterior distributions. 

Our best fit BLR model well reproduces both the differential phase and \brg{} line profile with a reduced chi-square, $\chi^2_{\mathrm{r}} = 0.665$. 
Fig.~\ref{fig:sc_visphi} shows the combined best fit BLR model added to the continuum phase signal for the baseline averaged differential phase spectra (see Fig.~\ref{fig:uvbin_bry_visphi} for the comparison to the $uv$-binned data). A representation of the on-sky cloud distribution is plotted in Fig.~\ref{fig:blr_model_circ}a and Fig.~\ref{fig:photocent}c shows the corresponding model photocentres which agree well with the orientation and size of the observed photocentres. We note that the PA of the model photocentres does not match the best fit PA. This is due to the moderate fraction of inflowing clouds which twists the PA on sky away from the PA of the rotating disk.

With a best fit inclination of $i = 23 \degree$ and $\theta_0 = 24 \degree$, we find a relatively face-on and moderately thick disk describes the BLR of NGC~3783 well, as expected for a type 1 AGN and similar to both 3C 273 \citep{GC2018Natur} and IRAS 09149-6206 \citepalias{Gravity-Collaboration:2020ab}. The centre of the BLR is offset from the centre of the hot dust by (-0.5, -19) \uas{} similar to the offset in the photocentres. In Fig.~\ref{fig:blr_model_circ}b we show the intrinsic BLR differential phase signal compared to the data by averaging the continuum phase subtracted differential phase spectra from the longest three baselines shown in Fig.~\ref{fig:blr_model_circ}c. Both the model and data show the characteristic ``S-shape'' expected of a BLR with kinematics dominated by Keplerian rotation. Fig.~\ref{fig:blr_model_circ}c also shows the decomposition of the model differential phases into the BLR and continuum components. Because we had already removed most of the continuum phase signal, the primary component is the BLR component. 

We find very little anisotropy in the emission of the BLR clouds. $\kappa$ is consistent with 0 indicating neither the near nor far side of the clouds is preferentially emitting. $\xi$ is $\sim0.7$ suggesting little mid-plane obscuration. Finally, $\gamma$ is close to 1.0 indicating the clouds on the outer surface of the BLR are not radiating more than the inner clouds. 

The motion of the clouds are split fairly evenly between bound elliptical orbits and radial motion with $f_{\rm ellip} \sim 0.46$. The clouds with radial motion are inflowing with $P_{\rm inflow} = 0.65$.\footnote{Instead of reporting the exact value of $f_{\rm flow}$ which has no physical meaning we instead report the probability that $f_{\rm flow} < 0.5$.} Our finding of a significant fraction of inflowing clouds agrees with the simple velocity resolved time lag measurements from \citet{Bentz:2020aa} which showed slightly shorter lags in the red wing of the line profile. 

An interesting feature of our best fit model is the radial distribution of the clouds. With $\beta=1.7$, which is moderately high, the distribution is more sharply peaked near $R_{\rm min}$ with a tail towards larger radii compared to 3C~273 and IRAS~09149-6206 which each had a $\beta\sim1.2$. The best fit distribution produces a mean radius, $R_{\rm BLR}=71$ \uas{} which corresponds to 16 light-days and a minimum radius, $R_{\rm min}=32$ \uas{} corresponding to 7 light-days. In our model, there is no maximum radius at which a cloud can exist. 

We can ask whether these various radii make physical sense. Reverberation mapping time lags for high ionization lines such as \ion{He}{ii}, \ion{C}{iv}, and \ion{Si}{iv} have been measured between 1--4 days \citep{Reichert:1994aa,Onken:2002aa,Bentz:2020aa} indicating gas at smaller radii than the $R_{\rm min}$ measured here. At large radii, both observations and models suggest a maximum radius of the BLR at the dust sublimation radius \citep{Laor:1993aa,Netzer:1993aa,Korista:1997aa,Baskin:2014aa,Schnorr-Muller:2016qy,Suganuma:2006aa} which for NGC~3783 should occur around 80 light-days. Our best fit model only contains $<1$\% of clouds at radii larger than this so indeed for NGC~3783 it does seem the model BLR is confined within the dust sublimation radius.

As a test of the sensitivity of the model to the inner and outer radii, we further decided to fit the data with a model with fixed $R_{\rm min} = 4$ light-days and $R_{\rm max} = 80$ light-days. We find that by fixing these radii we can still achieve a fit nearly as good as our fiducial one. The main change to the BLR model is that $\beta$ decreases to 1.2, $\kappa$ decreases to -0.3, and $f_\mathrm{ellip}$ increases to 0.83. All other parameters remain unchanged including \mbh{}. This highlights both the extreme flexibility of the BLR model and in particular the choice of the shifted gamma distribution to describe the radial distribution of the clouds as well as the insensitivity of the current data-set to $R_{\rm min}$. We will explore this issue in more detail in upcoming publications that will also include a joint reverberation mapping and GRAVITY analysis of NGC 3783 that will help to constrain these parameters. For the rest of this paper, we remain with the BLR parameters as presented in Table~\ref{tab:blr}.

In the following section we address and compare our mean radius with the time lags measured through reverberation mapping and discuss the SMBH mass and NGC~3783's place on the R-L relation.

\subsection{Time lags, the R-L relation, and SMBH mass}
Our BLR mean radius is a factor of 1.6 larger than the radius measured through reverberation mapping. However, in order to compare the model size from interferometry with reverberation mapping, we need to take into account that both techniques are sensitive to different parts of the cloud distribution in the BLR. Reverberation mapping determines a response-weighted mean radius while the interferometry modelling measures a deprojected brightness-weighted size of the distribution so differences between the two are not unexpected.

\begin{figure*}
    \centering
    \includegraphics[width=0.7\textwidth]{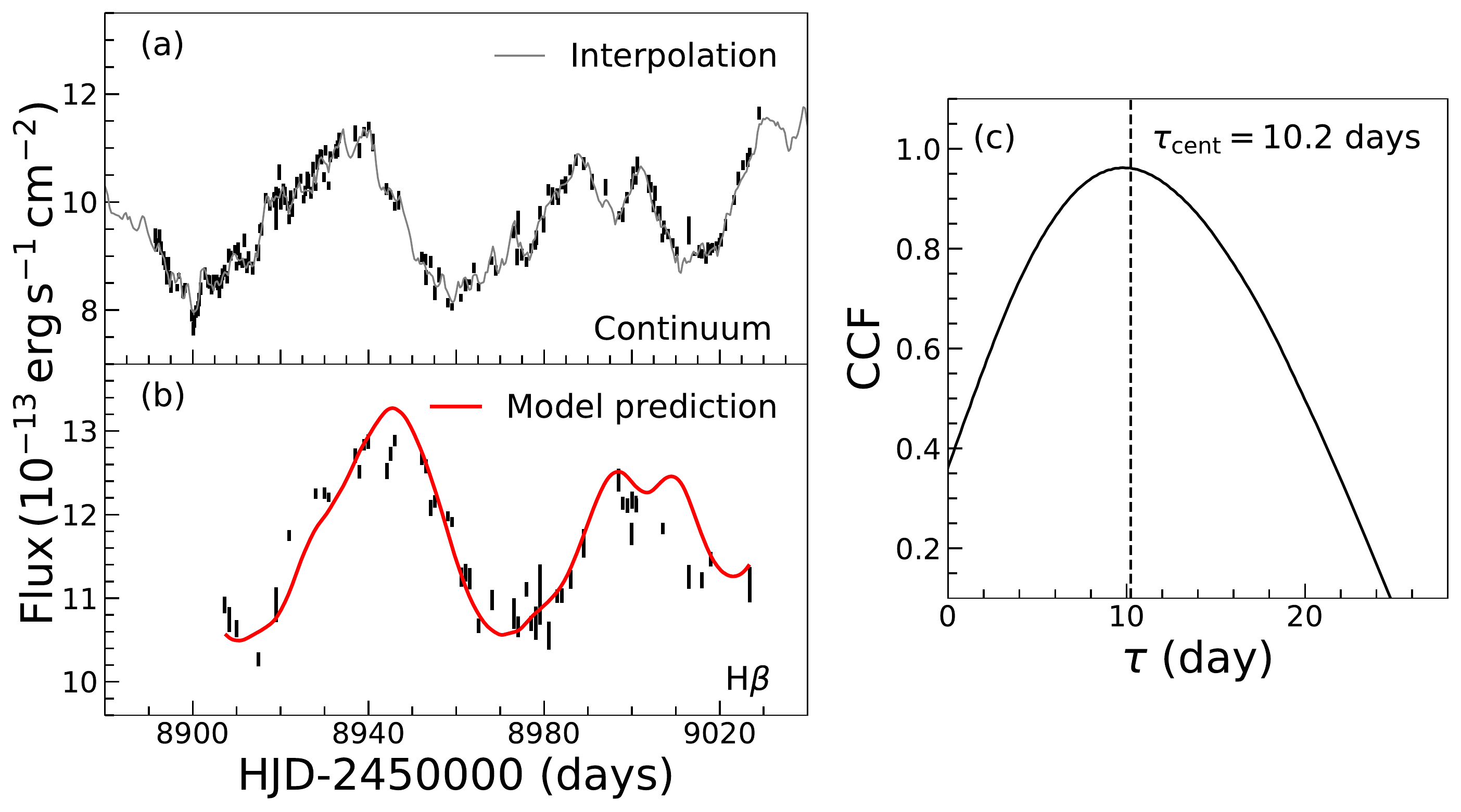}
    \caption{\textit{a)} Observed continuum light curve from \citet{Bentz:2020aa} (black points) with the finer sampled interpolated continuum light curve over-plotted (gray line). \textit{b)} Observed broad emission line light curve (black points) with the emission line light curve produced by reverberating the input continuum light curve off the clouds in our best fit BLR model (red). \textit{c)} Cross-correlation function between the continuum and broad emission line light curves as a function of time lag. The dashed line indicates the peak cross correlation time lag of 10.2 days.
    }
    \label{fig:reverberate}
\end{figure*}

To test this, we generated a mock emission line light curve based off the real continuum light curve from \citet{Bentz:2020aa}. 
From our best fit BLR model, we calculate a transfer function based on the distribution of time lags associated with each BLR cloud. This was then used to convolve the continuum light curve and produce an emission line light curve. Fig.~\ref{fig:reverberate}a shows the input continuum light curve and Fig.~\ref{fig:reverberate}b shows the emission line light curve corresponding to our best fit BLR model.
Following reverberation studies, we finally calculate the time lag between the two light curves using the cross correlation function (CCF). The results of the CCF are shown in Fig.~\ref{fig:reverberate}c.

\begin{figure*}
    \centering
    \includegraphics[width=0.7\textwidth]{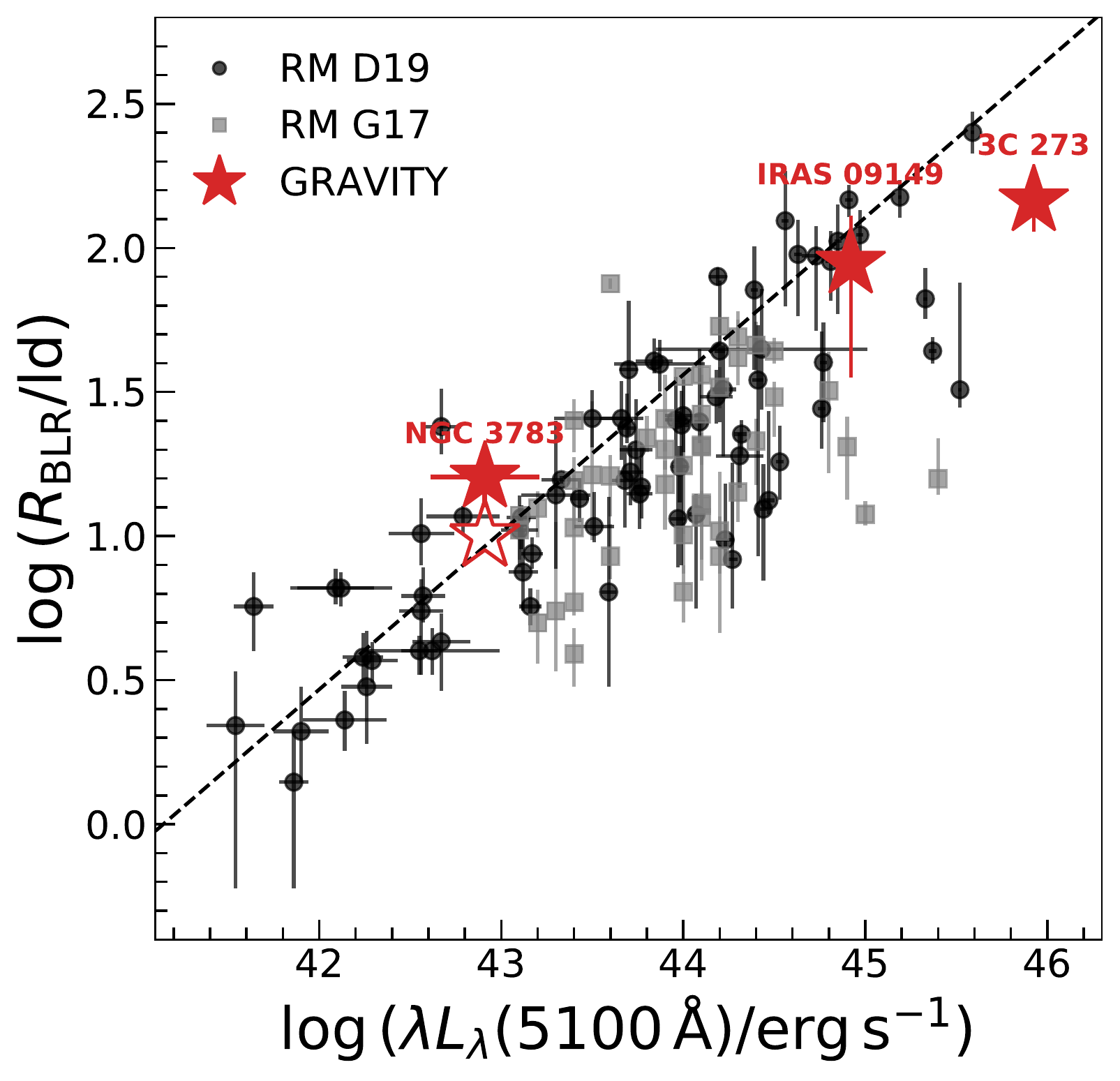}
    \caption{Relationship between BLR radius and 5100\AA{} AGN luminosity with the best fit line from \citet{Bentz:2013aa} (dashed line) and the RM samples from \citet{Du:2019aa} (black points) and \citet{Grier:2017aa} (grey points). Previous GRAVITY BLR radius measurements of IRAS 09149-6206 and 3C 273 are shown along with the BLR mean radius for NGC~3783 as a filled red stars. The time lag measurement of $\sim 10$ days for NGC~3783 from the literature is shown as an open star. The horizontal error bar indicates a factor of 2 uncertainty on the luminosity of NGC~3783 based on variability.}
    \label{fig:rl}
\end{figure*}

We find a peak time lag of $10.2^{+4.2}_{-2.7}$ days, very consistent with the time lags measured in previous studies. This suggests that, at least for NGC~3783, there is a modest difference between the physical mean radius of the BLR and the observed time lag. In Fig.~\ref{fig:rl}, we place NGC~3783 on the relation measured by \citet{Bentz:2013aa} using a 5100 \AA\ continuum luminosity, $\log\,\lambda L_{5100} = 42.93$ erg s$^{-1}$, from \citet{Bentz:2020aa} adjusted for our luminosity distance of 38.5 Mpc and both $R_{\rm BLR}$ determined from our modelling (filled star) and the measured time lag from the literature (open star). While the 5100 \AA\ continuum luminosity from \citet{Bentz:2020aa} was measured at about the same time as our last set of GRAVITY observations in March 2020, our full set of observations span several years. Long term monitoring of NGC~3783 shows that the AGN can vary by a factor of 2 over several years \citep{lira11}. Therefore, in Fig.~\ref{fig:rl} we include an uncertainty of 0.3 dex in the 5100 \AA\ continuum luminosity. GRAVITY-based results for IRAS 09149-6206 and 3C 273 are also plotted along with two RM samples from \citet{Du:2019aa} and \citet{Grier:2017aa}. We note here that both IRAS~09149-6206 and 3C~273 do not show the same difference between the interferometrically and RM measured size. This is due to the lower $\beta$ values for their cloud distributions and thus we only plot one point for them in Fig.~\ref{fig:rl}.

As seen before, the literature time lag for NGC~3783 lies perfectly on the $R-L$ relation \citep[e.g.][]{Peterson:2004aa,Bentz:2013aa}. However, the physical mean radius is above it by 0.2 dex. We note that NGC~3783's discrepancy with the $R-L$ relation is different than the increased scatter that both \citet{Du:2019aa} and \citet{Grier:2017aa} have reported. \citet{Du:2019aa} specifically showed that shorter time lags are observed in AGN with high Eddington ratios while \citet{Grier:2017aa} point to selection effects or a change in the BLR structure at higher luminosities rather than accretion rate. NGC~3783, on the other hand, is at lower luminosity and lower Eddington ratio ($\sim0.1$ based on the modelled black hole mass and $\log L_{\mathrm{bol}} = 44.52$ from \citepalias{Gravity-Collaboration:2020ac}) and we see a consistent time lag with the $R-L$ relation and rather find an offset mean radius. This highlights the fact that the $R-L$ relation is fundamentally a time-lag luminosity relation and the conversion to a radius from a time lag relies on simple assumptions about the geometry and structure of the BLR. \citet{White:1994aa} predicted that if the true BLR cloud distribution is significantly extended, reverberation mapping time lags could be biased towards the inner radius. 

The discrepancy we observe could also be related to the same trend we have observed with the dust sizes. \citepalias{Gravity-Collaboration:2020ac} showed that lower luminosity AGN have larger interferometric dust sizes compared to RM measured dust sizes. It is possible that both the BLR and hot dust geometry are changing in similar ways with increasing AGN luminosity. In particular, this could be related to the recent result that the obscuration fraction of AGN significantly decreases with increasing Eddington ratio due to the AGN driving gas and dust out of the central few parsecs \citep{Ricci:2017ek}. If the gas and dust is able to accumulate in the inner regions at low luminosity this would lead to the discrepancies we observe while at higher luminosity, gas and dust is driven out primarily from the inner regions and reduces the difference between RM and physical radii. This is certainly speculative and based primarily on two objects. We need more interferometric BLR measurements to test this explanation.

%**************************** R-L Relation Figure ************************************************************
%*************************************************************************************************************

Our BLR modelling also constrains the SMBH mass with $\log M_{\rm BH} = 7.68$. Interestingly, this is quite consistent with the masses based on the time lag and using a virial factor, $f \approx 4-5$, which is needed to match the $M-\sigma$ relation. Thus, the inferred virial factor for NGC~3783 is consistent with the average value used in RM studies.

\section{The structure of the hot dust continuum}\label{sec:hot_dust}
Beyond the BLR, we can also study, using the FT data, the surrounding hot dust structure which produces the underlying NIR continuum. Fig.~\ref{fig:ft_vis2_t3phi} shows a general trend of decreasing $V^2$ with increasing baseline length which indicates a partially resolved primary source of the hot dust continuum. \citetalias{Gravity-Collaboration:2020ac} used this data to measure a Gaussian FWHM size of 0.82 mas. However, \citetalias{Gravity-Collaboration:2020ac} also noted that NGC~3783 was the only AGN within their sample to show strong signatures of asymmetry evidenced by non-zero closure phases. The right panel of Fig.~\ref{fig:ft_vis2_t3phi} plots the closure phases which show consistently negative values between -2\degree{} and -10\degree. Symmetric structures such as a circular Gaussian cannot produce non-zero closure phases and thus higher-order structures must be present. In the following sections, we investigate the nature of the hot dust continuum using two methods 1) model independent image reconstruction and 2) $uv$ plane model fitting. 

\subsection{Continuum Image Reconstruction}\label{sec:image_reconstruct}

\begin{figure}
\centering
\includegraphics[width=\columnwidth]{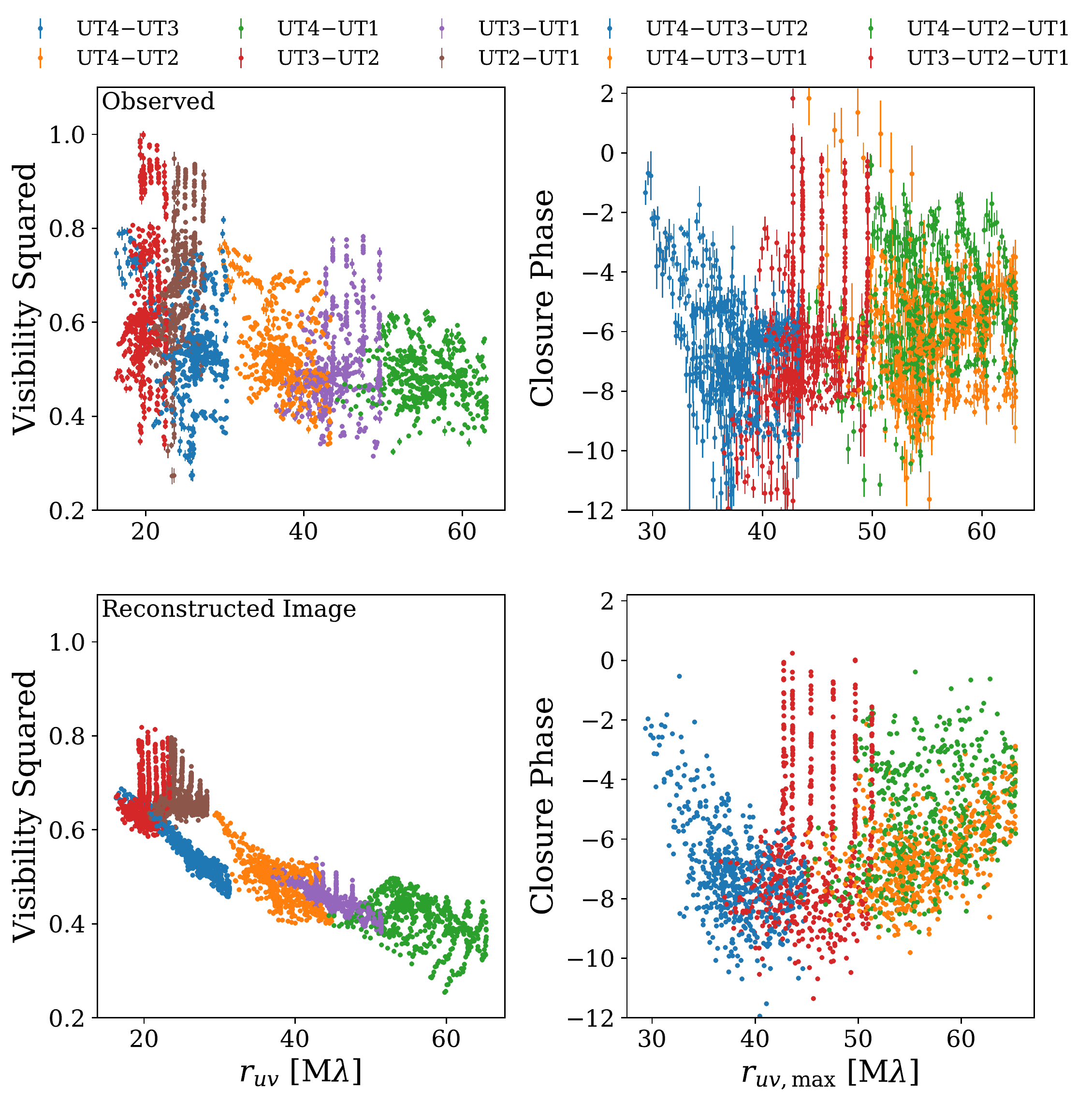}
\caption{GRAVITY FT squared visibilities (left) and closure phases (right) as a function of spatial frequency for all nights of observation. The significant non-zero closure phases indicate resolved asymmetric hot dust structure in NGC~3783.}
\label{fig:ft_vis2_t3phi}
\end{figure}

%**************MiRA Image Reconstruction Figure***********************
\begin{figure*}
\centering
\includegraphics[width=0.75\textwidth]{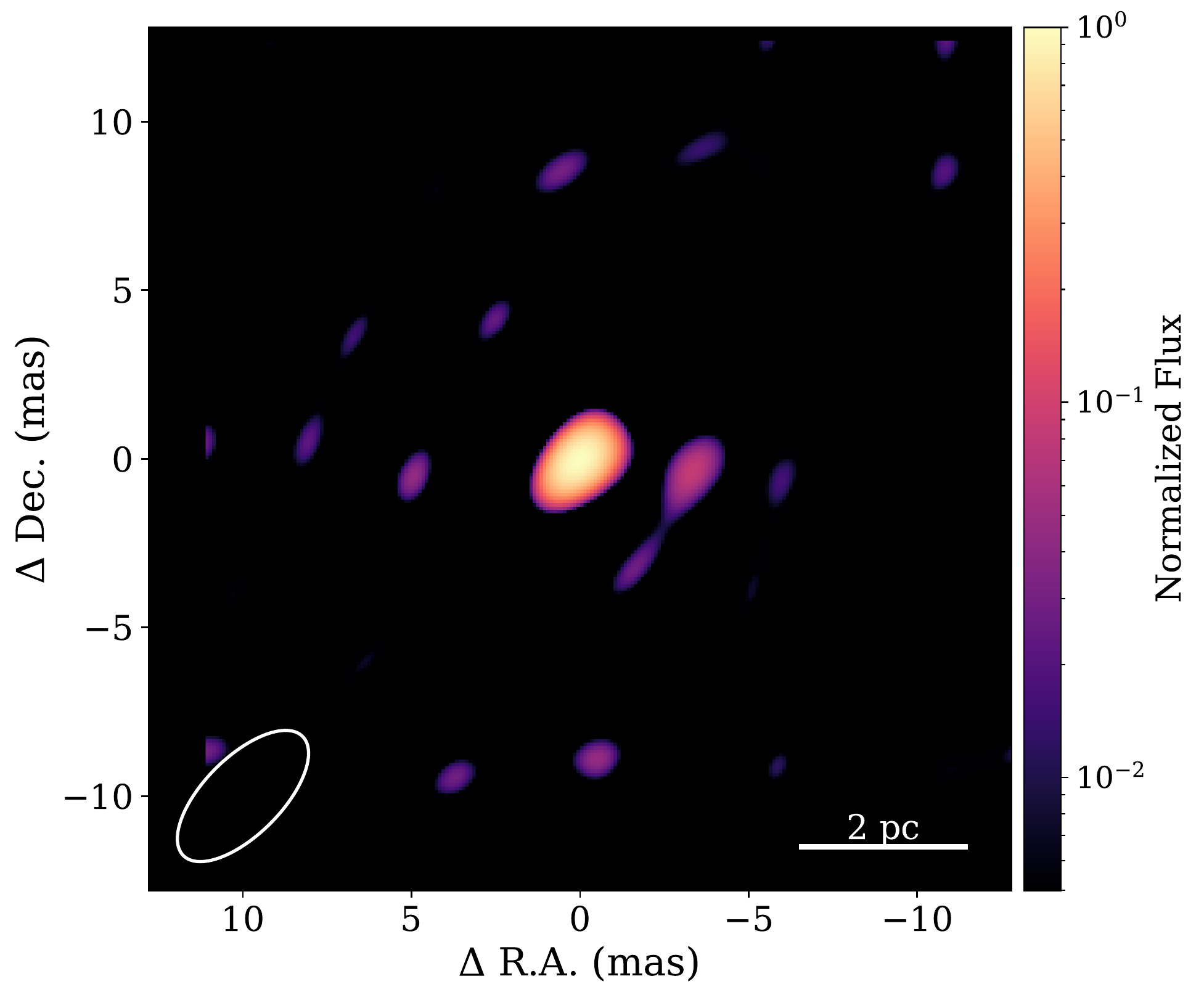}
\caption{\texttt{MiRA} image reconstruction of the K-band continuum for NGC~3783. We used the \textit{hyperbolic} regularisation with a FOV of 25.6 mas and a pixel size of 0.1 mas. The image is normalized to the maximum pixel flux. The GRAVITY interferometric beam is shown as a white ellipse in the bottom left corner for comparison. North is up and East is to the left. }
\label{fig:mira_image}
\end{figure*}

\begin{figure}
\centering
\includegraphics[width=\columnwidth]{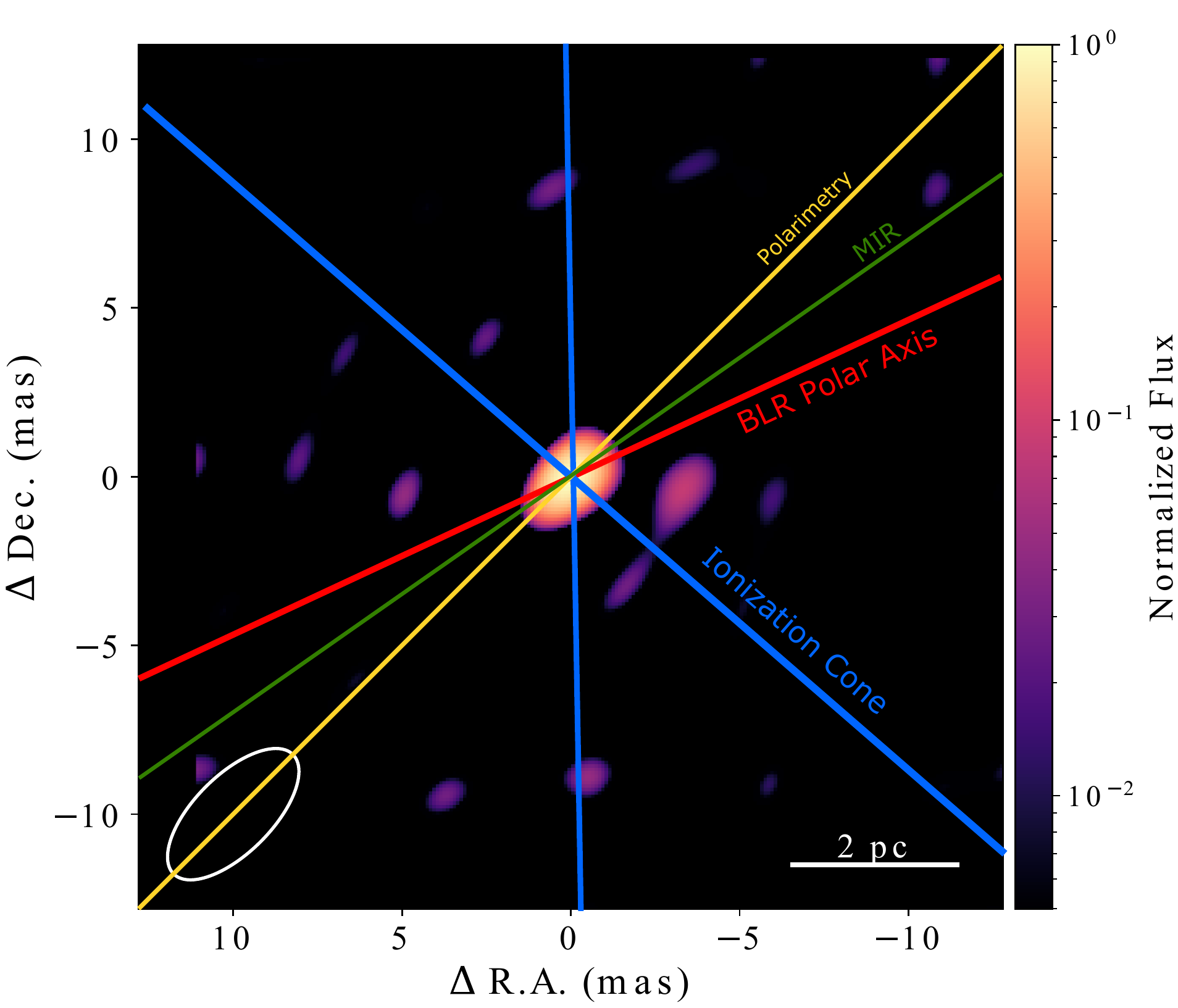}
\caption{Same as Fig.~\ref{fig:mira_image} but with lines indicating important angles and regions. Blue lines outline the ionization cone as produced by the BLR polar axis (red line) and thickness. Yellow line indicates the position angle from polarimetery \citep{lira20} and the green line indicates the position angle from MIR interferometry \citep{Honig:2013qy}.}
\label{fig:mira_image_lines}
\end{figure}

%*********************************************************************

The high quality and large $uv$ coverage of the FT data allows us to utilise image reconstruction codes to spatially map the hot dust. As we did for our imaging of NGC 1068 \citep{Gravity-Collaboration:2020aa}, we used \texttt{MiRA}\footnote{publicly available at \url{https://github.com/emmt/MiRA}} \citep[Multi-aperture Image Reconstruction Algorithm][]{Thiebaut:2008aa} to reconstruct the K-band image of NGC~3783 based on the $V^{2}$ and closure phase data as shown in Fig.~\ref{fig:ft_vis2_t3phi}. The $V^{2}$ uncertainties shown have been multiplied by 10 to match the S/N of the closure phases and avoid overweighting $V^2$. This also takes into account calibration uncertainty that is not included in the pipeline generated data.

In general, image reconstruction from interferometric data is an ill-posed problem due to sparse coverage of the $uv$ plane, especially for optical/NIR interferometry. Therefore, to reduce the number of possible solutions (i.e. images) that can equally reproduce the data, image reconstruction codes use priors to constrain the brightness distribution. Within \texttt{MiRA} we imposed both a positivity prior (i.e. flux must be $\geq0$) and the \textit{hyperbolic} regularisation which is an edge-preserving smoothness prior. The \textit{hyperbolic} regularisation favours solutions where the flux is smooth inside the structure but contains sharp edges. Two hyperparameters can be tuned with this regularisation, $\mu$ and $\tau$. $\mu$ is simply the weight given to the regularisation in determining the total likelihood, $L = f_{\rm data} + \mu f_{\rm prior}$, where $f_{\rm data}$ is a function comparing the model to the data and $f_{\rm prior}$ is a function comparing the model to the constraints given by the prior.

$\tau$ is a hyperparameter specifically associated with the \textit{hyperbolic} regularisation and is an edge threshold that controls how sharp the edges are expected to be. Too small values of $\tau$ will produce a cartoon-like image with very smooth regions that are then sharply cutoff. Large values of $\tau$ instead lead to effectively a \textit{compactness} regularisation which favours a single centrally concentrated source. 

\citet{Thiebaut:2017aa} outline best practices for choosing optimum values of $\mu$ and $\tau$. Following these, we ran \texttt{MiRA} over a grid of values for $\mu$ and $\tau$. For each value of $\tau$, we chose the value of $\mu$ that corresponds to the elbow of the ``L-curve'' which is a plot of $f_{\rm data}$ against $f_{\rm prior}$. Choosing $\mu$ at the elbow is a compromise between over- and under-regularisation. We then visually inspected all of the images associated with each $\mu$-optimised $\tau$ value and chose the image that avoided the cartoon-like effects but also many spurious compact sources. We found optimum values of $\mu$ and $\tau$ of $5\times10
^5$ and $10^{-3}$ respectively. 

Fig.~\ref{fig:mira_image} shows our final image reconstruction at a pixel scale of 0.1 mas, a FOV of 25.6 mas, and an initial image of a Dirac delta function . As expected, the hot dust image is dominated by a central, marginally resolved source, however immediately noticeable is the presence of a fainter smaller source to the SW. Simple Gaussian fitting directly on the image finds a size of 1.8x1.2 mas and PA $\sim-44$\degree{} (East of North) for the bright central source and 1.5x0.8 mas and PA $\sim-32$\degree{} for the fainter SW source (hereafter referred to as the ``offset cloud''). The offset cloud is 3.3 mas (0.6 pc)  away from the centre at a PA of $\sim-96$\degree{} and contains 5\% of the flux. 

The PA and extent of the individual sources follow the PA of the GRAVITY beam therefore it is possible these properties are an artefact of the reconstruction and rather reflect the limited $uv$ coverage of our data. Importantly, the distance and direction of the offset source places it well outside the beam of the central source and strengthens the reliability of its detection. In Appendix~\ref{app:mira_tests}, we show that the offset source is robust against the choice of regularisation, choice of image reconstruction algorithm, and random removal of data, while the other much fainter sources are likely spurious. 

The bottom panels of Fig.~\ref{fig:ft_vis2_t3phi} show the $V^2$ and closure phases of our image reconstruction. The closure phases match the observed ones quite well and accurately reproduce all of the key features including the rise to 0\degree{} towards smaller spatial frequencies and the vertical streaks prominent in the UT3$-$UT2$-$UT1 triangle. $V^2$, however, is only moderately well matched. In particular, the gradient of $V^2$ seems steeper for the image reconstruction which leads to a larger primary source. This leads to the factor of $\sim2$ difference between the image reconstructed size and the size found in \citetalias{Gravity-Collaboration:2020ac} (0.82 mas) from fitting a Gaussian model to only $V^2$ data. Therefore, as a further test of the detection and specific properties of each source, in the next section we apply the same model fitting.

\subsection{Visibility model fitting}

We fit the individual $V^2$ and closure phase data of each night with a model composed of a central 2D Gaussian, an offset point source, and an unresolved background. As for the image reconstruction, we inflated the $V^2$ errors by a factor of 10 to match the S/N of the closure phases and partially account for calibration uncertainty. 

\begin{figure*}
\centering
\begin{tabular}{cc}
\includegraphics[width=0.45\textwidth]{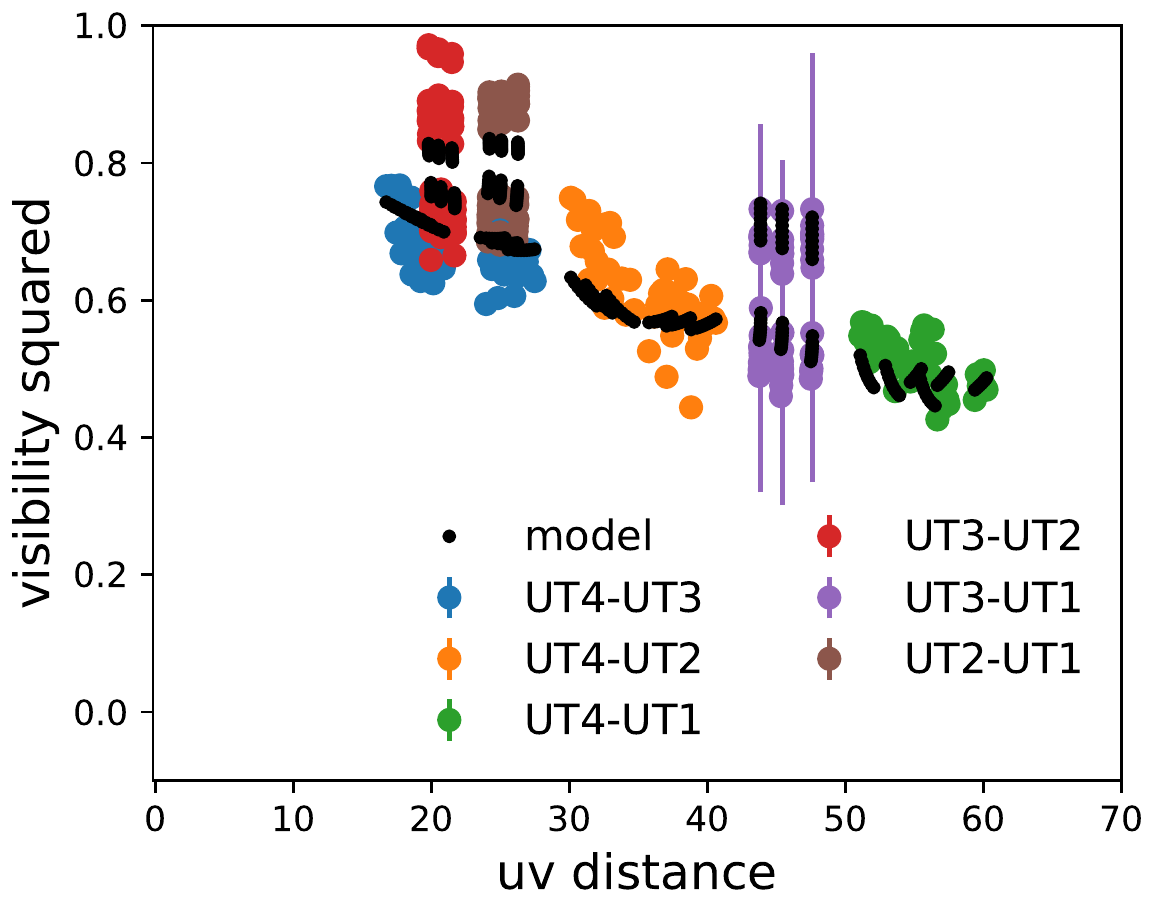} &
\includegraphics[width=0.45\textwidth]{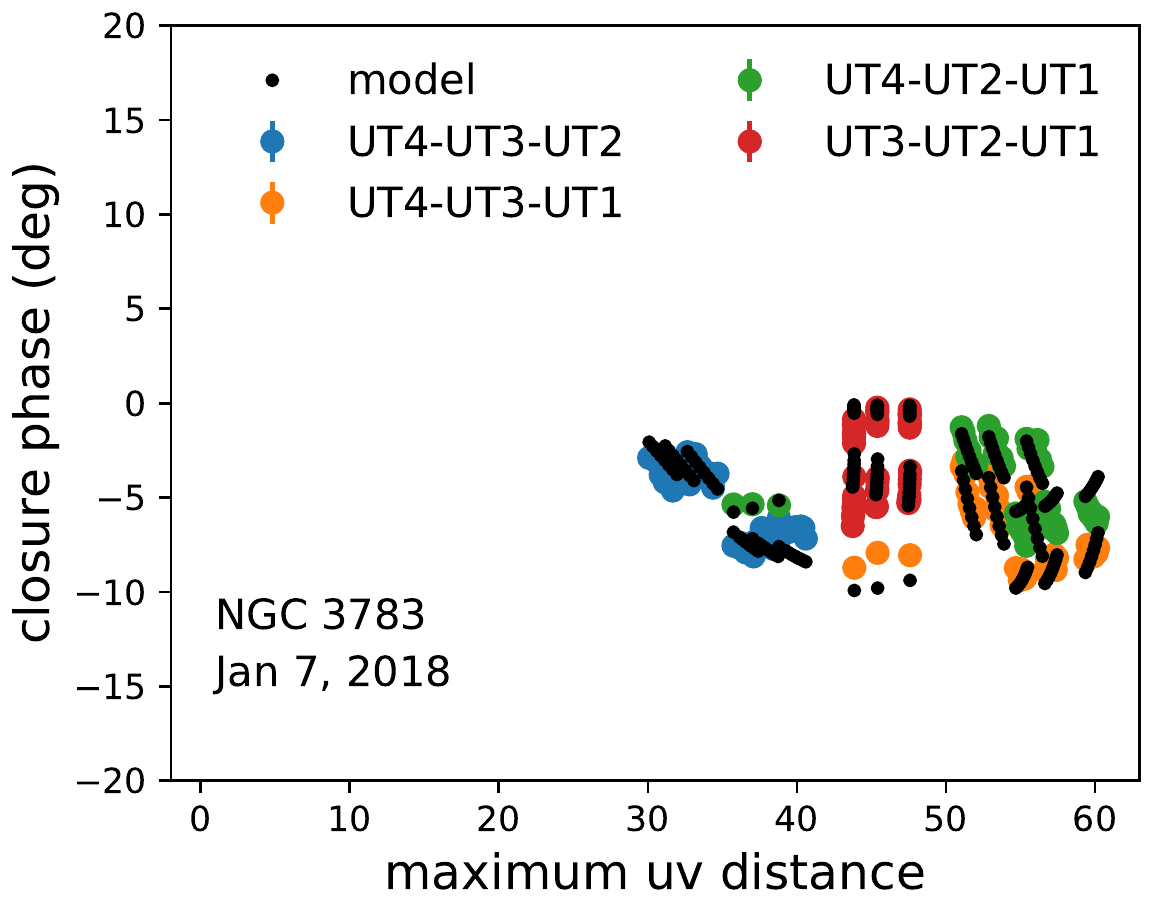}
\end{tabular}
\caption{Sample 2D Gaussian plus point source model fit to the observed $V^2$ and closure phase data for NGC~3783 from Jan 7, 2018. The best fitting Gaussian FWHM geometric mean size of $\simeq 0.6$ mas is consistent with that found from 1D Gaussian fitting \citepalias{Gravity-Collaboration:2020ac}. The Gaussian PA and point source flux fraction and offset are consistent with those found from image reconstruction.}
\label{fig:dust_modelfit}
\end{figure*}

Table~\ref{tab:model_fits} lists the best fit parameters for each night and Fig.~\ref{fig:dust_modelfit} shows an example fit for January 7, 2018. In all nights, we find good qualitative agreement with the features of the data, and in particular, the offset point source provides a good match to the observed non-zero closure phases as was found in the image reconstruction. Between nights, we also find very good consistency in the best fit results with an average central source FWHM of $0.72 \pm 0.1$ mas, and an offset point source at ($-2.9 \pm 0.2$, $-1.0 \pm 0.2$) mas relative to the central Gaussian emitting an average fractional flux of $0.043 \pm 0.01$.  This places the offset point source 3.1 mas (0.57 pc) away from the central Gaussian at a PA of $-109$\degree{}. All values except for the size of the central Gaussian are in excellent agreement with the features seen in our reconstructed image.  

While our model fitting agrees well with \citetalias{Gravity-Collaboration:2020ac}, we suspect the choice of regularisation is causing the much larger size in the reconstructed image. Indeed, both increasing $\tau$ and switching to a \textit{compactness} regularisation reduced the size of the central Gaussian to values similar to those in Table~\ref{tab:model_fits}. This illustrates how sensitive image reconstruction can be on the choice of regularisation especially since we are at sizes well below the diffraction limit, $\Delta \theta \ll \lambda/B$. For this reason, we choose to use the average Gaussian FWHM found through our model fitting as the size of the main hot dust continuum source for this paper. Both methods, image reconstruction and visibility model fitting, robustly detect the offset cloud and we discuss possible origins in the next section.  

%*********** Model Fitting Results Table ***************************
\input{tab4}
%*******************************************************************

\subsection{The origin of the offset hot dust source}

We explore two possibilities for the origin of the offset dust cloud: 1) an orbiting secondary SMBH and 2) a dust cloud heated by the central AGN. 
In both cases, we use a projected radial distance from the primary AGN of 0.6 pc and a K-band flux density of 3.3 mJy. This flux density was calculated starting with a total K-band fibre magnitude of 10 mag \citepalias[][]{Gravity-Collaboration:2020ac} and using the measured $\sim5$\% fractional flux of the offset cloud.  

In scenario 1, we could be observing a secondary SMBH orbiting the primary AGN. The hot dust emission would then be indicative that the secondary SMBH is also a faint AGN that is heating the dust around it and would also contain its own BLR. The secondary would then have a NIR luminosity of $10^{42}$ erg s$^{-1}$ and a bolometric luminosity of $10^{42.7}$ erg s$^{-1}$ using the NIR-X-ray relation from \citet{Burtscher:2015lr} and converting to a bolometric luminosity using the relation in \citet{Winter:2012yq}. Assuming the secondary AGN is at the Eddington luminosity places a lower limit of the black hole mass of $4 \times 10^4$ M$_{\odot}$ and assuming it is accreting at 10\% Eddington, similar to the primary, gives \mbh{} $= 4\times10^5$ M$_{\odot}$. As a first order calculation of the expected orbital velocity of the secondary, we can simply use $v = \sqrt{GM/r}$ with $M = 10^{8}$ M$_{\odot}$ found from our BLR modelling since the secondary must be much less massive than the primary. At a radius of 0.6 pc, we would then expect $v \approx 900$ km s$^{-1}$ which would induce a shift in the secondary BLR and produce a double peak in the combined spectrum, albeit at only 5\% the strength of the primary broad emission lines and assuming there is little obscuration towards the secondary BLR. This velocity is also the maximum velocity we would expect given a fully inclined orbit and the secondary currently at its maximum projected distance from the primary. We attempted to fit the normalised \brg{} profiles with the addition of a second broad emission component but could not achieve any reasonable fit. Another potential signal of a binary SMBH would be a periodic light curve, however given the distance of the secondary and the mass of primary, the expected period of the orbit is 4400 years. Thus, the current data on NGC~3783 cannot fully rule out a SMBH binary as the origin of the offset cloud.

A simpler explanation is scenario 2, where the offset cloud is a massive cloud of gas either inflowing towards the AGN from the circumnuclear disk or potentially outflowing away from it. The dust emission then would be heated, not internally, but externally by the central AGN. Indeed as shown in Fig.~\ref{fig:mira_image_lines}, the cloud does lie near the edge of the ionization cone that is produced given the geometry of our best fit BLR which further matches the geometry inferred from polarimetry \citep{lira20} and MIR interferometry \citep{Honig:2013qy}. This would give the cloud a direct view towards the AGN and its radiation.

To test this scenario, we can use the observed luminosity of the AGN to constrain the dust temperature and measure a dust mass under the assumption that it is emitting as a modified blackbody. If an nonphysical amount of dust would be needed to produce the NIR emission observed then we can safely rule out this scenario. We calculate the dust temperature using the simple scaling relation that the temperature decreases as a power law with a maximum dust temperature, $T_{\rm sub}$ at the dust sublimation radius, $r_{\rm sub}$, \citep[e.g.][]{Honig:2010lr},

\begin{equation}\label{eq:dust_temp}
\frac{r}{r_{\rm sub}} = \left(\frac{T}{T_{\rm sub}}\right)^{\alpha}
\end{equation}

\noindent Both $\alpha$ and $r_{\rm sub}$ depend on the specific grain distribution. We use the ``ISM large grains'' model from \citet{Honig:2010lr} which consists of 47\% graphites and 53\% silicates and a \citet{Mathis:1977aa} size distribution between 0.1 \micron{} and 1 \micron{}. For this dust model, $r_{\rm sub} = 0.5\,\rm{pc}$ for $T_{\rm sub}=1500\,\rm{K}$ and an AGN bolometric luminosity of $10^{46}$ erg s$^{-1}$ and $\alpha = -2.1$. We adjust for NGC~3783's AGN luminosity ($\log\,L_{\rm bol} = 44.5$) by scaling with $L_{\rm bol}^{1/2}$ and find $r_{\rm sub} = 0.09$ pc, consistent with the radius measured for the bright component (0.07 pc) in our visibility model fitting. At the location of the offset cloud, we then calculate a dust temperature of 604 K. We note that other grain models, such as those based on pure graphite grains in the innermost part of the obscuring structure, give similar $r_\mathrm{sub}$.

With only one spectral energy distribution (SED) data point, we use a simple modified blackbody model to calculate the dust mass.

\begin{equation}\label{eq:mod_bb}
S_{\nu} = \frac{M_{\mathrm{d}}\kappa_{0}}{D_{\mathrm{L}}^2}\left(\frac{\nu}{\nu_{0}}\right)^{\beta}\frac{2h\nu^{3}}{c^{2}}\frac{1}{e^{{h\nu/kT_{\mathrm{d}}}}-1}\,,
\end{equation}

\noindent where $M_{\mathrm{d}}$ is the dust mass, $D_{\mathrm{L}}$ is the luminosity distance, $c$ is the speed of light, $h$ is the Planck constant, and $k$ is the Boltzmann constant. We use $\kappa_{0}=167\,\mathrm{m^{2}\,kg^{-1}}$, $\beta=1.25$, and $\nu_{0}=137\,\mathrm{THz}$ (2.19 \micron) from \citet{Draine:2003gd}. With $T_{\mathrm{d}} = 604$ K from Eq.~\ref{eq:dust_temp} and $S_{K} = 3.3$ mJy, we estimate $M_{\mathrm{d}} = 0.2$ M$_{\odot}$. Assuming a standard dust-to-gas ratio of 100, then the gas mass is 20 M$_{\odot}$. This cloud mass does not seem completely unreasonable, especially on sub-pc scales around an AGN. Assuming a radius of $\sim0.2$ pc for the cloud directly measured from the image, the density of the cloud would be $\sim10^4$ cm$^{-3}$ which is at the high end of the observed densities in the NLR of local AGN \citep{Davies:2020aa}. Therefore, while we cannot specifically dismiss the binary SMBH origin of the offset cloud, we prefer the primary AGN heated dust scenario for simplicity.

MATISSE, the new mid-IR instrument at the VLTI could potentially help distinguish between our two proposed explanations. While we showed that we expect a dust temperature $T_{\rm d} \sim 600$ K if heated by a single central AGN, if instead it is heated by a secondary it should have a more complex SED, likely with an inner hot dust component at $\sim1400$ K and cooler dust at further radii \citep[e.g.][]{Kishimoto:2011aa}. Using the simple modified blackbody, we estimate flux densities of 26, 45, and 28 mJy for the L, M, and N bands accessible by MATISSE. Based on the exposure time calculator, closure phase uncertainties of 1\degree{} should be achievable with a few hours of observations in the L and M bands. 

\section{The nuclear size of the Coronal Line Region}\label{sec:clr}

\begin{figure*}
\centering
\includegraphics[width=0.7\textwidth]{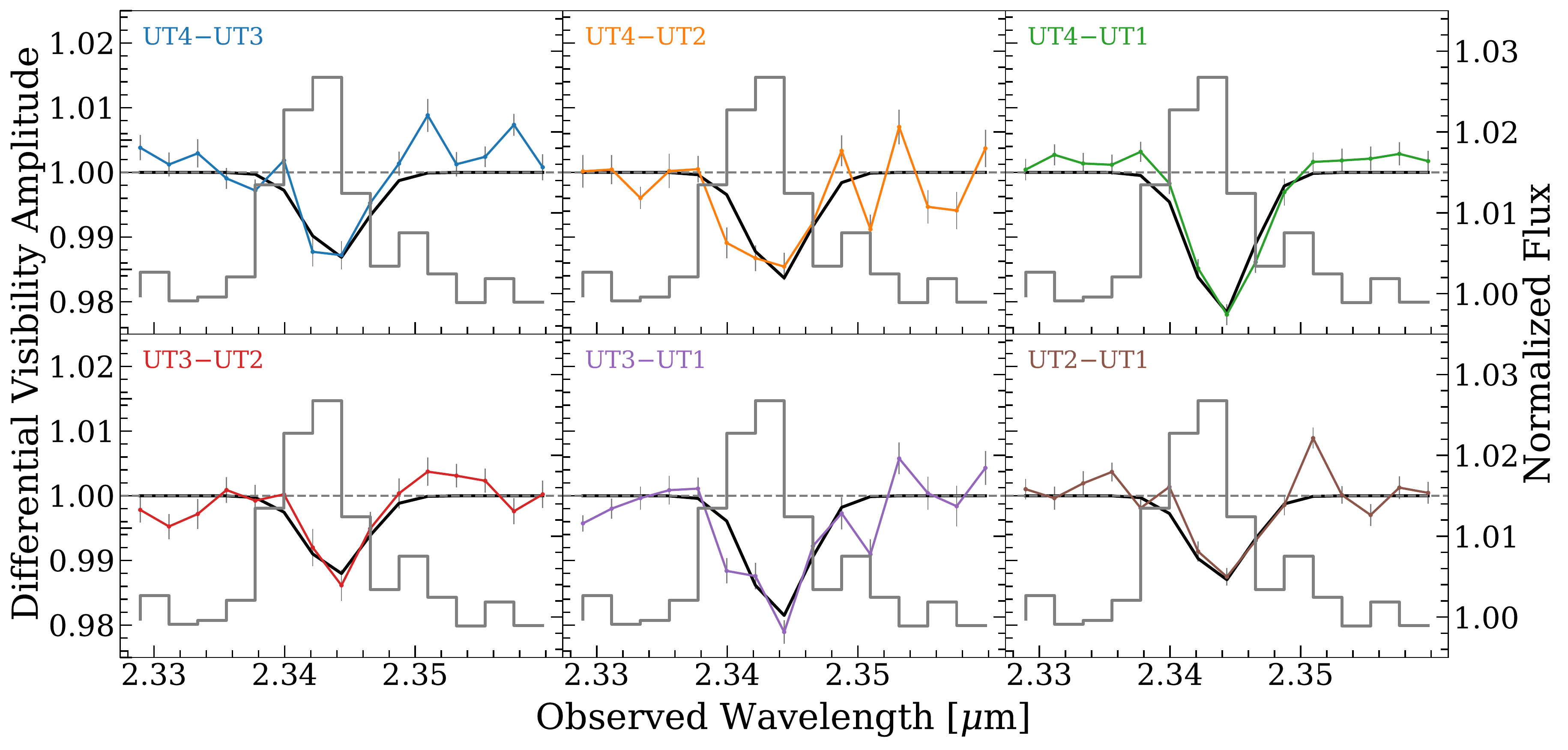}
\caption{Same as Fig.~\ref{fig:sc_visamp} but for the \caviii{} region.}
\label{fig:sc_visamp_caviii}
\end{figure*}

Our final analysis from this rich data set involves, for the first time, measuring the size of the coronal line region (CLR) on nuclear scales. In addition to the BLR and NLR, a subset of AGN also show lines from highly ionized atoms, the so-called ``coronal lines'', for their first observation in the solar corona. These lines all have ionization potentials $\geq 100$ eV and are collisionally excited forbidden transitions with relatively high critical densities ($10^7 - 10^{10}$ cm$^{-3}$). Because of the high critical densities and observations that many coronal lines have FWHM intermediate between BLR and NLR lines \citep[e.g][]{Appenzeller:1991aa,Veilleux:1991aa,Rodriguez-Ardila:2006aa,Rodriguez-Ardila:2011aa,Lamperti:2017aa} it is thought that the CLR lies between the BLR and NLR. A distinct increase in line FWHM with ionization potential for some AGN is further evidence of its intermediate location \citep[e.g][]{Wilson:1979aa,Penston:1984aa,Thompson:1995aa,Rodriguez-Ardila:2002aa,Rodriguez-Ardila:2011aa} under the assumption of photoionization as the primary ionization mechanism \citep[e.g][]{Penston:1984aa,Ferguson:1997aa,Mazzalay:2010aa,Rodriguez-Ardila:2011aa}. This has also led to the hypothesis that the CLR originates at the inner region of the central obscuring structure \citep{Pier:1995aa,Ferguson:1997aa,Murayama:1998aa}. Arguments against this hypothesis though include the fact that coronal lines are observed equally in type 1 and type 2 AGN \citep[e.g.][]{Rodriguez-Ardila:2011aa} and that significant coronal line emission is seen in even the most heavily Compton thick AGN like Circinus, Centaurus A, and NGC 1068 \citep[e.g.][]{Moorwood:1996aa,Reunanen:2003aa,Rodriguez-Ardila:2006aa,Mazzalay:2010aa,Muller-Sanchez:2011aa}.

Further, through long-slit and integral field spectroscopic observations, coronal line emission has also been found to be extended on $\sim$10--100 pc scales \citep[e.g][]{Prieto:2005ab,Mazzalay:2010aa,Muller-Sanchez:2011aa}, but with the brightest emission still concentrated in the unresolved nuclear region. This more extended and more diffuse emission instead is likely to be powered by shocks that could be produced as radio jets propagate through the surrounding gas \citep[e.g.][]{Rodriguez-Ardila:2002aa,Reunanen:2003aa,Rodriguez-Ardila:2006aa,Muller-Sanchez:2011aa,Rodriguez-Ardila:2017aa}. GRAVITY provides the first opportunity to measure the size of the nuclear CLR through the \caviii{} line.

The differential visibility amplitude spectra spanning the \caviii{} line (Fig.~\ref{fig:sc_visamp_caviii}) show a dip near the peak of the emission line in all the baselines, in contrast to the \brg{} spectra which show primarily a peak and indicated a smaller BLR size compared to the hot dust. A dip instead is caused by a larger emitting line region compared to the hot dust. We also observe a dip in differential visibility amplitude superimposed on the broad peak seen in the \brg{} region, and the location of this dip further corresponds to the narrow peak in the flux profile which suggests that the narrow \brg{} line emitting region is also larger than the hot dust continuum. The narrow \brg{} line then could be tracing the base of the NLR or also originate in the CLR. 

The differential visibility amplitude across an emission line is described by the following equation,

\begin{equation}\label{eq:diff_amp_single}
\Delta V = 
\frac{1 + f V_\mathrm{line}/V_c}{1 + f},
\end{equation}

\noindent where $f$ is the flux line profile normalised to a continuum of 1, $V_\mathrm{line}$ is the visibility amplitude of the line, and $V_c$ is the visibility amplitude of the continuum. \brg{}, however, is the combination of a broad and narrow line, each with a different line width and physical size relative to the continuum. The differential visibility amplitude for the \brg{} line then follows,

\begin{equation}\label{eq:diff_amp_double}
\Delta V_{Br\gamma} = 
\frac{1 + f_{b} V_b/V_c + f_{n} V_n/V_c}{1 + f_{t}},
\end{equation}

\noindent where $f_{t}$, $f_{b}$, and $f_{n}$ are the total, broad, and narrow flux profiles normalised to a continuum of 1, and $V_{c}$, $V_{b}$, and $V_{n}$ are the continuum, broad, and narrow visibility amplitudes. 

To fit both Eq.~\ref{eq:diff_amp_single} and \ref{eq:diff_amp_double} we model all emitting regions as Gaussian sources. In the marginally resolved limit, the visibility of a Gaussian follows,

\begin{equation}\label{eq:vis_gauss}
    V = V_0\mathrm{exp}\left(\frac{-\pi^2 r_{uv}^2 \mathrm{FWHM}^2}{4\log2}\right)
\end{equation}

\noindent where $V_0$ is the zero baseline visibility, $r_{uv}$ is the baseline length in units of mas$^{-1}$, and FWHM is the size of the source in mas. While a single compact source should always have $V_0=1$, \citetalias{Gravity-Collaboration:2020ac} showed this is not the case for our AGN and can be caused either by coherence loss or unresolved background emission. We therefore choose to include this in our modelling. We further set $V_b=1$ since the BLR is effectively unresolved. The measured size of 101 \uas{} from spectroastrometry (see Section~\ref{sec:blr_model}) results in $V_b = 0.995$ at the longest baseline observed. 

Eq.~\ref{eq:diff_amp_single} and \ref{eq:diff_amp_double} also show that the differential visibility amplitude depends on the normalised flux profile. While for the \caviii{} line we could simply use the observed line profile, for \brg{} we need to decompose the line into its narrow and broad components. To fold in the uncertainties related to this decomposition, we include in our model the shape of the narrow and broad components, parameterised as Gaussian lines with an amplitude, central wavelength, and FWHM. We further also fit for the FWHM Gaussian size of the narrow \brg{} emitting region. In total our model has 14 free parameters to fully describe the sizes of the CLR, hot dust, and narrow \brg{} regions; the differential visibility amplitude spectra across the \brg{} and \caviii{} lines; and the flux line profiles of \brg{} and \caviii{}.

%****************************************************
% Table for best fit model parameters of CLR fitting
\input{tab3}
%****************************************************

Table~\ref{tab:clr} lists the fitted parameters and their best fit values and uncertainties that were estimated using the median and 95\% credible interval of their respective posterior distributions (see Fig.~\ref{fig:corner_clr}). From this analysis, we measure a hot dust size of 0.73 mas corresponding to 0.13 pc and a CLR size of 2.2 mas corresponding to 0.4 pc. The hot dust size very well matches the size measured from visibility model fitting of the FT data (See Section~\ref{sec:hot_dust}). 

Our size firmly places the nuclear \caviii{} emitting clouds beyond both the BLR and the NIR emitting hot dust. For NGC~3783, this argues against an origin of \caviii{} in the inner region of the central obscuring structure which has been suggested as the site of the CLR \citep{Pier:1995aa,Ferguson:1997aa,Murayama:1998aa}. This however does not rule out an inner region origin for all coronal lines especially since \caviii{} has one of the lowest ionisation potentials of the observed coronal lines (IP = 128 eV). Indeed, \citet{Mullaney:2009aa} were able to model the flux and kinematics of multiple coronal lines with an AGN driven outflow launched from inner edge of a dusty obscuring structure. The high velocity components of the highest ionisation lines all are produced near the dust sublimation radius while the lower ionisation lines are produced at larger radii. This could also explain the low velocity shift of the \caviii{} and relatively low FWHM since at larger radii, the gravitational potential of the stellar bulge should slow down the clouds.

The \caviii{} flux\footnote{We calculated the \caviii{} flux from the GRAVITY spectrum and assuming a constant continuum flux of 10 mag.} of $4\times10^{-18}$ W m$^{-2}$ observed from a region of radius 0.92 mas can be used to derive a crude estimate for the density of this inner part of the CLR. We adopt an estimated filling factor by \caviii{} emitting gas $f=0.1$, to exclude regions either devoid of gas or with Ca in other ionisation stages. Using atomic data of \citet{Landi:2004aa} and \citet{Saraph:1996aa} for a simple 2-level analysis, and a Ca abundance of 2.e-6 by number \citep{Landi:2004aa}, $n_e\approx 2\times10^5$ cm$^{-3}$ is needed to reproduce the observed \caviii{} flux. This is consistent with the notion that coronal lines arise in an inner and dense part of the narrow line region, see, e.g., \citet{Davies:2020aa} for a recent assessment of NLR densities. The required CLR density scales with $f^{-0.5}$, i.e. the electron density in the \caviii{} emitting region must be at least about $10^5$ cm$^{-3}$ but could be clearly higher if arising in very low filling factor clouds or filaments.

\section{Connecting the nuclear and circumnuclear regions}\label{sec:discuss}
With these new GRAVITY observations of NGC~3783, we have detected and measured the properties of three distinct components within a radius of 1 pc from the AGN:

\begin{enumerate}
    \item A rotating BLR with a mean radius of 0.013 pc, an inclination of 23\degree{}, and PA of 295\degree{}.
    \item A hot dust structure composed of a central bright source with a 0.14 pc size and a faint offset cloud 0.6 pc away at a PA of -109\degree{}.
    \item A nuclear CLR with a size of 0.4 pc.
\end{enumerate}

In this section, we aim to place these components in the context of the larger scale circumnuclear environment that has been well studied with previous observations. To help in this, we fit the \sivi{} and the ro-vibrational H$_{2}$ (1--0) S(1) lines detected in our SINFONI cube. We chose to also fit \sivi{} because it is a brighter coronal line than \caviii{} and has a similar ionization potential (IP = 167 eV) and thus allows for tracing the ionized gas out to larger scales. The H$_2$ line traces hot molecular gas that is likely inflowing and feeding the AGN \citep[e.g][]{Hicks:2009aa,Hicks:2013ly,Davies:2014aa}. 

Each emission line was fit with a single Gaussian profile on top of a linear continuum to trace the bulk motion of the line emitting gas. Spectral regions around each line were chosen to avoid other lines and regions strongly affected by telluric features (1.965--1.995 \micron{} for \sivi{} and 2.128--2.157 for H$_{2}$). Only spaxels which had at least one spectral channel with S/N$>3$ were fit where the noise was determined as the local line-free RMS of the spectrum. For \sivi, we masked the 1.975--1.978 \micron{} region where the H$_{2}$ (1--0) S(3) line is expected. Velocities were allowed to be $\pm1000$ km s$^{-1}$ and velocity dispersions were allowed to be 0--500 km s$^{-1}$. Finally, 100 Monte Carlo iterations of the fit were performed by adding Gaussian noise to the spectra to determine the uncertainties on the best fit line parameters.

\begin{figure*}
    \centering
    \includegraphics[width=0.8\textwidth]{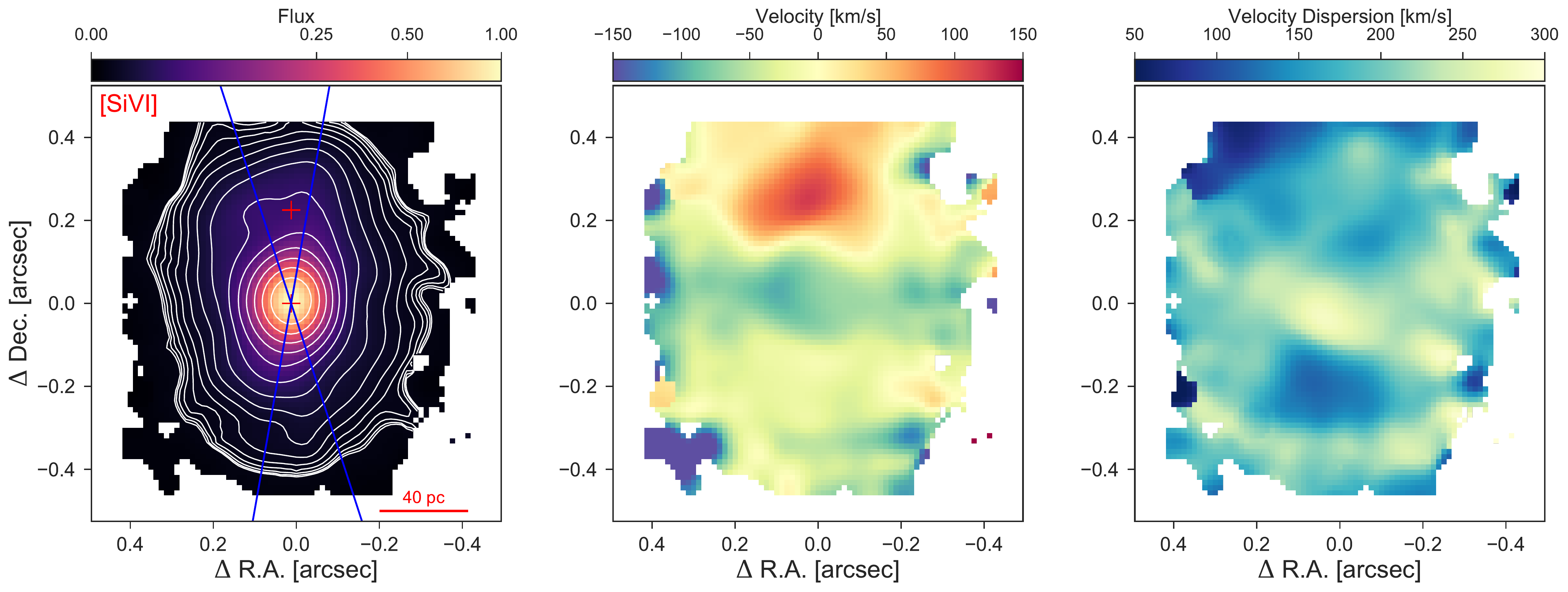}
    \includegraphics[width=0.8\textwidth]{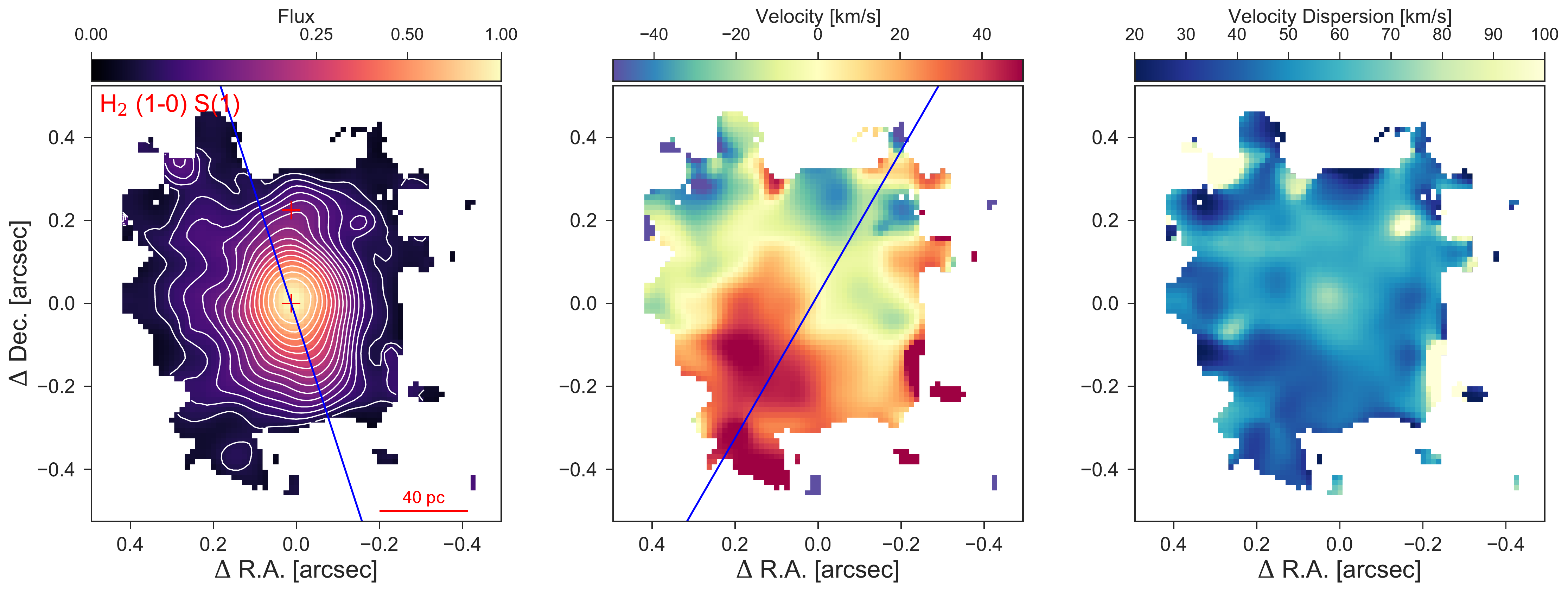}
    \caption{Normalised flux (left), LOS velocity (middle), and velocity dispersion (right) maps of the \sivi{} line (top row) and H$_{2}$ (1--0) S(1) line (bottom row). The maps were measured from the same SINFONI cube from which we derived the \brg{} profile used in our BLR modelling. The red crosses indicate the locations of the spaxels used to show example line fits in Fig.~\ref{fig:example_fits} The blue lines in the \sivi{} flux map show a PA of -18\degree{} and +10\degree{} which were estimated from the contours of the flux distribution (white contours). The blue line in the H$_{2}$ flux map shows a PA of +10\degree{} estimated from the contours of the flux distribution (white contours). The blue line in the H$_{2}$ velocity map shows a PA of -20\degree{} estimated visually from the velocity field.}
    \label{fig:sinfo_line_maps}
\end{figure*}

\begin{figure}
    \centering
    \includegraphics[width=\columnwidth]{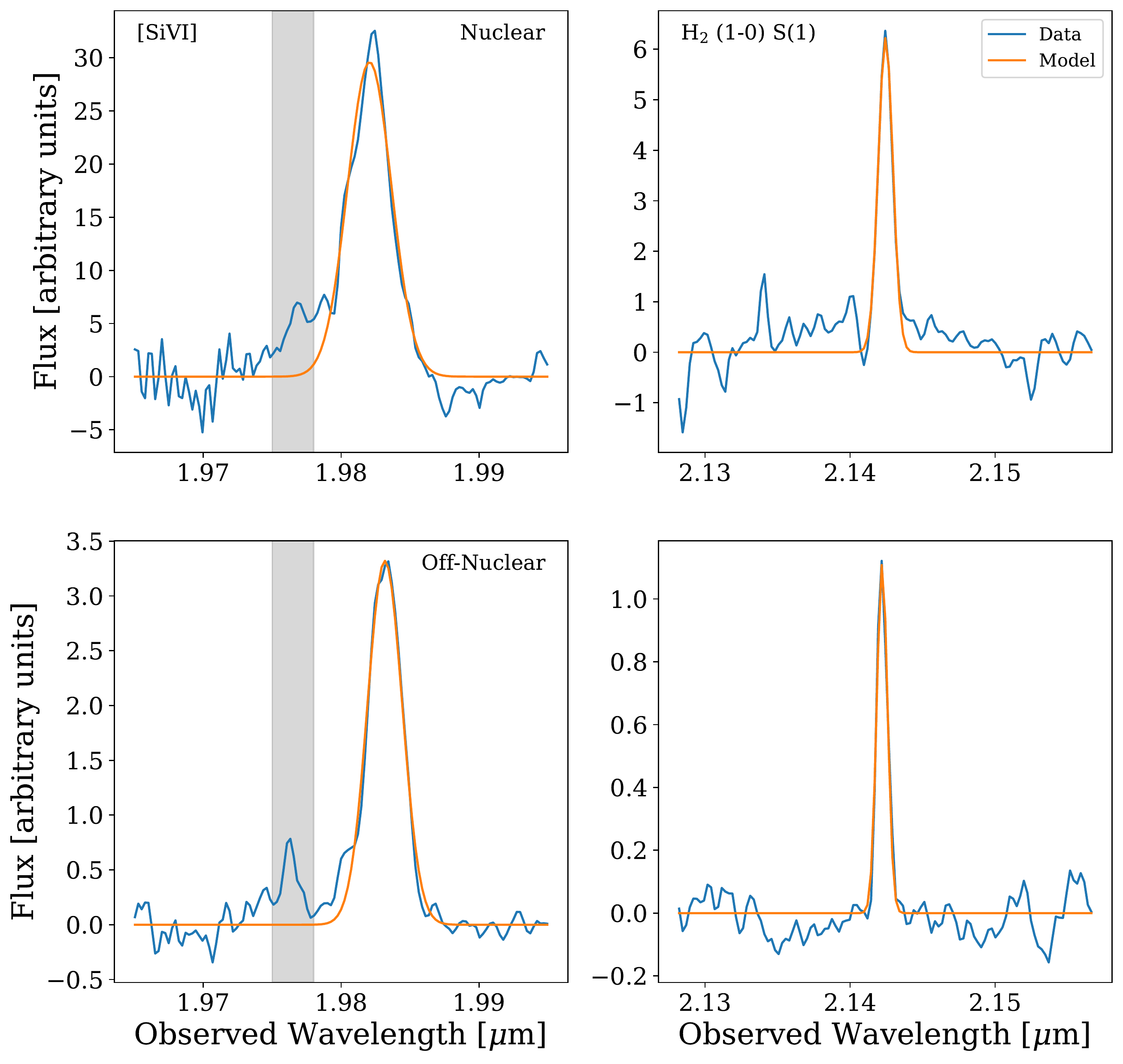}
    \caption{Example fits to individual spectra for the \sivi{} (left panels) and H$_{2}$ (1--0) S(1) line (right panels). The observed spectra are shown in blue while the best fit Gaussian models are shown in orange. The top row shows the spectral regions from a nuclear pixel while the bottom row shows the spectral regions from an off-nuclear pixel. The specific pixels used are shown as white crosses in Fig.~\ref{fig:sinfo_line_maps}. The grey shaded region in the \sivi{} panels show the masked region corresponding to the expected location of the H$_2$ (1--0) S(3).}
    \label{fig:example_fits}
\end{figure}

Fig.~\ref{fig:sinfo_line_maps} shows the results of our fits where velocities have been corrected for the systemic velocity of the host galaxy as given in NED (2917 km s$^{-1}$). In addition we also show example fits to two pixels in Fig.~\ref{fig:example_fits}. The location of the pixels are plotted as red crosses in Fig.~\ref{fig:sinfo_line_maps}. There is a clear difference in the structure and kinematics of the ionised and molecular gas. The flux distribution of \sivi{} seems to change PA from $\sim -18$\degree{} on small scales to $\sim +10$\degree{} on large scales. These PAs were measured by visually inspecting the contours shown in the \sivi{} flux map and therefore only represent estimates with uncertainties of 5\degree. Lines representing the PAs are also shown in the \sivi{} flux map as blue lines. 

The \sivi{} kinematics show clear non-circular signatures with a strong redshifted component to the North and high velocity dispersion. This matches the analysis of \citet{Muller-Sanchez:2011aa} who interpreted and modelled the kinematics as an outflow with a small contribution from disk rotation. The hot molecular gas, in contrast, shows a flux distribution with a PA of $\sim +10$\degree{} on all scales which is shown as a blue line in the H$_2$ flux map of Fig.~\ref{fig:sinfo_line_maps}. The kinematics are more indicative of disk rotation with a kinematic major axis along a PA of $\sim-20$\degree{} (shown as a blue line in the  H$_2$ velocity map of Fig.~\ref{fig:sinfo_line_maps} and estimated visually from the gradient of the velocity field) matching previous SINFONI results \citep{Davies:2007kx,Hicks:2009aa,Muller-Sanchez:2011aa}. The LOS velocities are also quite low ($\pm50$ km s$^{-1}$) and the dispersions relatively high (for a rotating disk) suggesting a low inclination, thick disk.

\begin{figure*}
\centering
\includegraphics[width=0.6\textwidth]{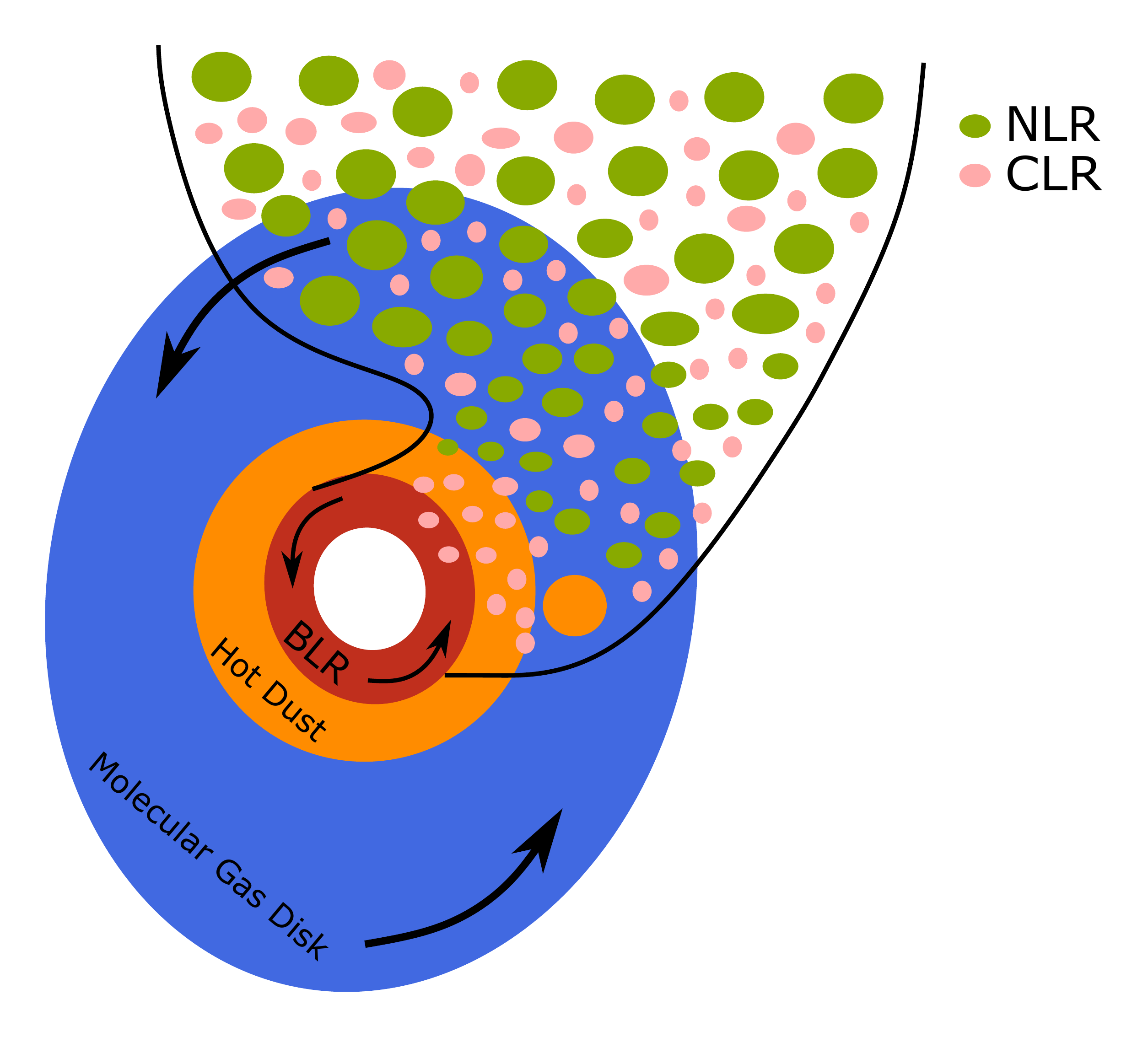}
\caption{A cartoon of the nuclear and circumnuclear region as described in Sec.~\ref{sec:discuss}. Different coloured clouds within the ionization cone correspond to coronal and narrow emission line emitting clouds. Arrows for the BLR and molecular gas disk indicate the direction of rotation. The image is not to scale in order to be able to show all components together.}
\label{fig:cartoon}
\end{figure*}

Interestingly, neither the \sivi{} nor the hot H$_2$ axes match the kinematic axis of the BLR. The larger scale flux distributions are close but the kinematic axes of the \sivi{} and H$_2$ are very different from the BLR. The BLR is blueshifted to the northeast and redshifted to the southwest while \sivi{} is redshifted to the north and blueshifted to the south. H$_2$ is blueshifted to the north but much more to the northwest. The hot molecular gas disk has a measured inclination of $\sim35$\degree{} \citep{Davies:2007aa,Hicks:2009aa,Muller-Sanchez:2011aa} which is twice the inclination of the BLR but matches the larger kpc scale inclination of the host galaxy. Thus, there must be a warping of the gas disk as it flows from 50 pc down to sub-pc scales. Such a warping has been observed in other AGN, most notably NGC 1068 \citep{Impellizzeri:2019aa} which shows counter-rotation at larger scales compared to the sub-pc maser disk.

Our BLR orientation however is in relatively good agreement with the polar axis measured by \citet{smith02} and \citet{smith04}. We measure a BLR polar axis of -65\degree{} compared with a polar axis of -45\degree{} measured through polarization. While this is a difference of 20\degree{}, the uncertainty on the BLR polar axis is large (+55\degree, -49\degree). This further matches the polar dust angle of -50 to -60\degree{} found through mid-IR interferometry \citep{Honig:2013qy,Burtscher:2013aa,lopezgonzaga16}. Our BLR inclination also well matches the inferred inclination from the disk+wind model of \citet{hoenig17} lending more evidence in favour of the interpretation that the extended MIR emission is tracing a dusty outflow.

To connect to the larger scale outflow traced by \sivi{}, there must be a gradual shifting of the orientation as the outflow has expanded and interacted with the host galaxy ISM. On parsec scales the outflow begins with a PA of $\sim-60$\degree{}, in line with the polar axis of the BLR and the extended MIR component. The outflow then seems to shift northward first to $-18$\degree{} and finally $+10$\degree{} matching the host galaxy disk. We note this is all still consistent with the \caviii{} size measurement which is only probing the very nuclear regions. Coronal line emission is easily produced at small scales due to their high critical densities and thus its expected for the brightest emission to occur in the nucleus. At larger scales, where the density is lower, it is likely that shock excitation instead of photoionization is producing the coronal lines and is expected to be fainter \citep{Rodriguez-Ardila:2017ab,May:2018aa}. Indeed the \sivi{} map shows a dominant unresolved central core and fainter, more diffuse extended emission up to 100 pc. To quantitatively determine the excitation mechanism at every scale would require detailed line modelling which is out of the scope of this Paper. However, \citet{Rodriguez-Ardila:2006aa} was able to explain the nuclear Fe and Si coronal line emission of NGC 3783 with photoionization alone but required additional shock excitation at radii larger than 100 pc. These shocks could be produced by a radio jet interacting with the interstellar medium of NGC 3783, however both VLA and VLBA observations show only an unresolved component at scales less than $\sim$20 pc \citep{Schmitt:2001aa,Orienti:2010aa}. More likely is that the shocks are produced by the AGN radiation pressure driven wind.

An interesting thing to note is we do not observe a large velocity shift in the nuclear \caviii{} line while at 55 pc we observe a LOS velocity of 150 km/s. Assuming a constant inclination angle, this could indicate an acceleration of the ionized gas from sub-pc to 10s of pc scale as seen in other AGN \citep[e.g][]{Muller-Sanchez:2011aa}. Another explanation could be that the outflow was launched with a distribution of velocities such that the material with the largest velocities is at the largest radii which has been observed in the NLR of NGC~1068 \citep{Miyauchi:2020aa}. 

\section{Summary}
In this paper, we reported on our analysis of VLTI/GRAVITY observations of the nearby type 1 AGN NGC~3783. We investigated three distinct components of the nuclear region around the AGN: 1) the broad line region, 2) the hot dust, and 3) the coronal line region. Our main conclusions are as follows:

\begin{itemize}
    \item We detect and successfully model the BLR as a rotating, thick disk with a physical mean radius of 16 light days. Due to the centrally peaked but heavy tailed distribution of clouds, this leads to a cross-correlation function measured peak time lag of $\sim10.2$ light days, fully consistent with reverberation mapping results. We find that the key parameters of interest for this Paper like \mbh{} appear to be robustly determined but others, in particular $R_\mathrm{min}$, are not. In a future paper exploring joint modelling of GRAVITY and reverberation mapping data we will return to this issue.
    
    \item We reconstruct an image of the hot dust which reveals the presence of an offset cloud of gas and dust 0.6 pc in projected distance away from the main central hot dust component. We measure a FWHM size of 0.14 pc for the main component consistent with previous interferometric results and with the expected dust sublimation radius.
    
    \item We measure a gas mass for the offset cloud of 20 M$_{\odot}$ and interpret it primarily as an AGN heated cloud outflowing within the ionization cone.
    
    \item We measure a FWHM size for the nuclear \caviii{} emitting CLR of 0.4 pc, firmly placing it at the very inner regions of the NLR. Combined with our VLT/SINFONI data, we show the CLR is composed of a bright compact nuclear component and a fainter extended component out to 100 pc with outflow kinematics.
    
    \item We combine our results with past MIR interferometric and our VLT/SINFONI integral field unit data to establish a comprehensive view of the nuclear and circumnuclear region of NGC~3783 that includes an extended dusty outflow originating along the polar axis of the BLR. The AGN sits within a thick molecular gas disk that is feeding the AGN. Both the outflow or the molecular gas disk could be the origin of the offset cloud seen in the image reconstruction.

\end{itemize}

In Fig.~\ref{fig:cartoon} we show all of the components together in a single picture in an effort to place them all into context with each other. We note the cartoon is not to scale so all inferred distances and sizes of components are not correct. We have further inferred a counterclockwise rotation direction for the gas given the winding direction of the larger scale spiral arms of NGC 3783 (see \citet{den-Brok:2020aa} for a recent image) which in most spiral galaxies trail the direction of rotation \citep[e.g.][]{Buta:2011aa}.  However, the relative orientations do match the description described here. Putting all of these observations together produces a comprehensive and coherent picture of the gas structure and dynamics from 0.01 -- 100 pc around an AGN that was only achievable with the impressive capabilities of NIR interferometry and VLTI/GRAVITY. With GRAVITY, we aim to perform a similar analysis for the brightest and nearest AGN, but the upgrade to GRAVITY+ will open a window to fainter and higher redshift AGN to be able to trace the evolution of gas around AGN as a function of cosmic time.

\begin{acknowledgements}
J.D. was supported in part by 
NSF grant AST 1909711 and an Alfred P. Sloan Research Fellowship.  A.A. and P.G. were supported by Funda\c{c}\~{a}o 
para a Ci\^{e}ncia e a Tecnologia, with grants reference UIDB/00099/2020 and 
SFRH/BSAB/142940/2018.  SH acknowledges support from the European 
Research Council via Starting Grant ERC-StG-677117 DUST-IN-THE-WIND. This research has made use 
of the NASA/IPAC Extragalactic Database (NED), which is operated by the Jet 
Propulsion Laboratory, California Institute of Technology, under contract with 
the National Aeronautics and Space Administration. This research made use of \textsc{Astropy},\footnote{http://www.astropy.org} a community-developed core Python package for Astronomy \citep{Astropy:2013ek,Astropy:2018dk}, \textsc{numpy} \citep{vanderWalt:2011we}, \textsc{scipy} \citep{Jones:2001ch}, and \textsc{matplotlib} \citep{Hunter:2007}
\end{acknowledgements}

%*******************************
% References
\bibliographystyle{aa}
\bibliography{my_bib.bib}
%*******************************

% Appendices

\appendix

\section{$uv$-binned differential spectra and model fits}\label{app:uv_bin_spec}
In this Appendix, we show the $uv$-binned differential visibility phase and amplitude spectra along with the best fit BLR and CLR models described in the main text. Also shown are the full joint and marginalised posterior distributions for our BLR and CLR fits.
\begin{figure*}
    \centering
    \includegraphics[width=0.8\textwidth]{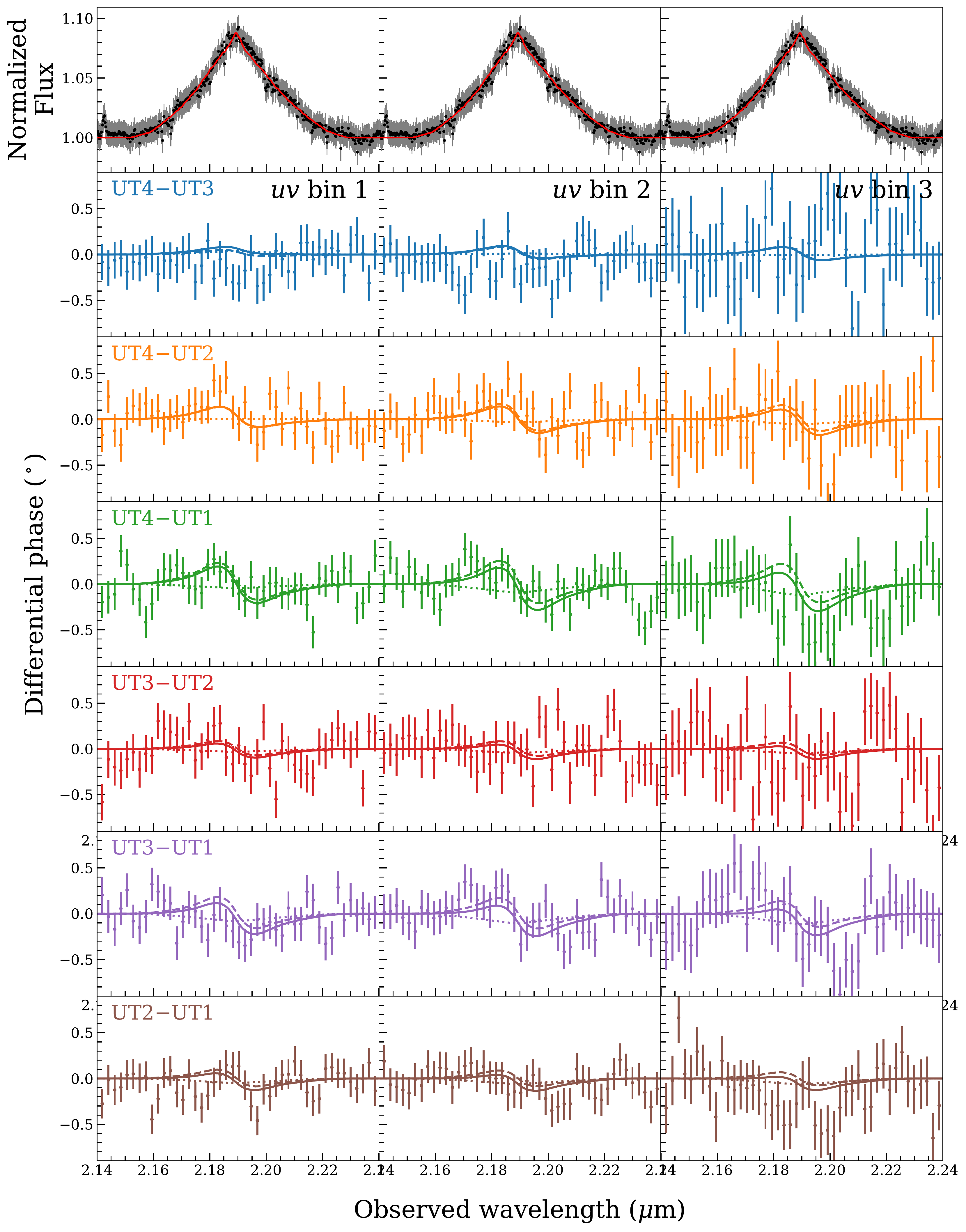}
    \caption{\textit{top row:} Normalized \brg{} profile (black points) with best fit model line profile (red line). \textit{bottom rows:}$uv$-binned, continuum phase subtracted differential phase spectra for each baseline (coloured points) with the best fit BLR model spectra (solid lines). Model spectra for the BLR component are shown as dashed lines while the offset component is shown as dotted lines.}
    \label{fig:uvbin_bry_visphi}
\end{figure*}
\begin{figure*}
    \centering
    \includegraphics[width=\textwidth]{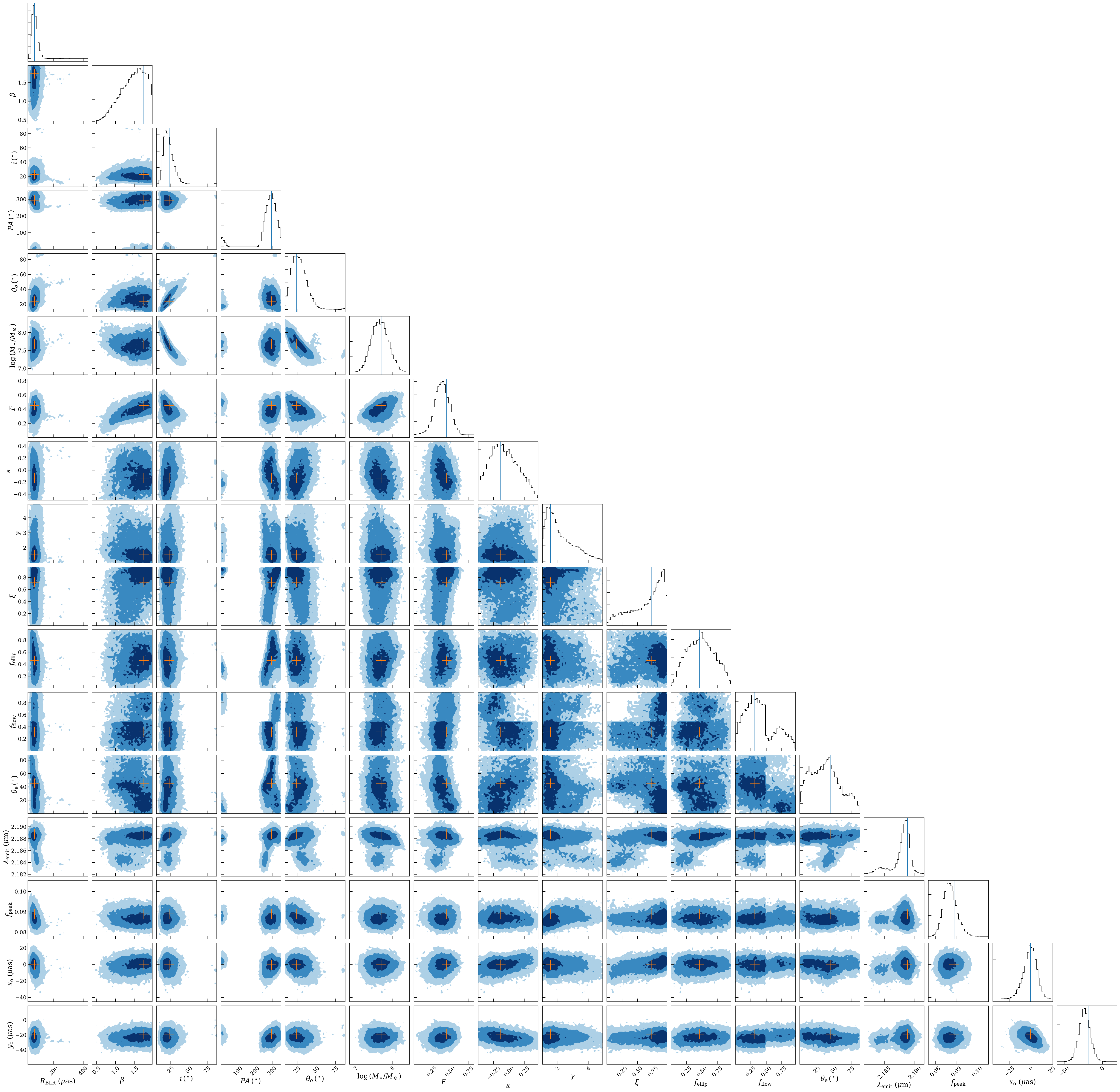}
    \caption{Corner plot for our BLR model fit showing the joint and marginalised posterior distributions for each free parameter. The blue lines and orange crosses indicate the maximum \textit{a posteriori} position used as the best fit values given in Table~\ref{tab:blr}}
    \label{fig:corner_blr}
\end{figure*}
\begin{figure*}
    \centering
    \includegraphics[width=0.8\textwidth]{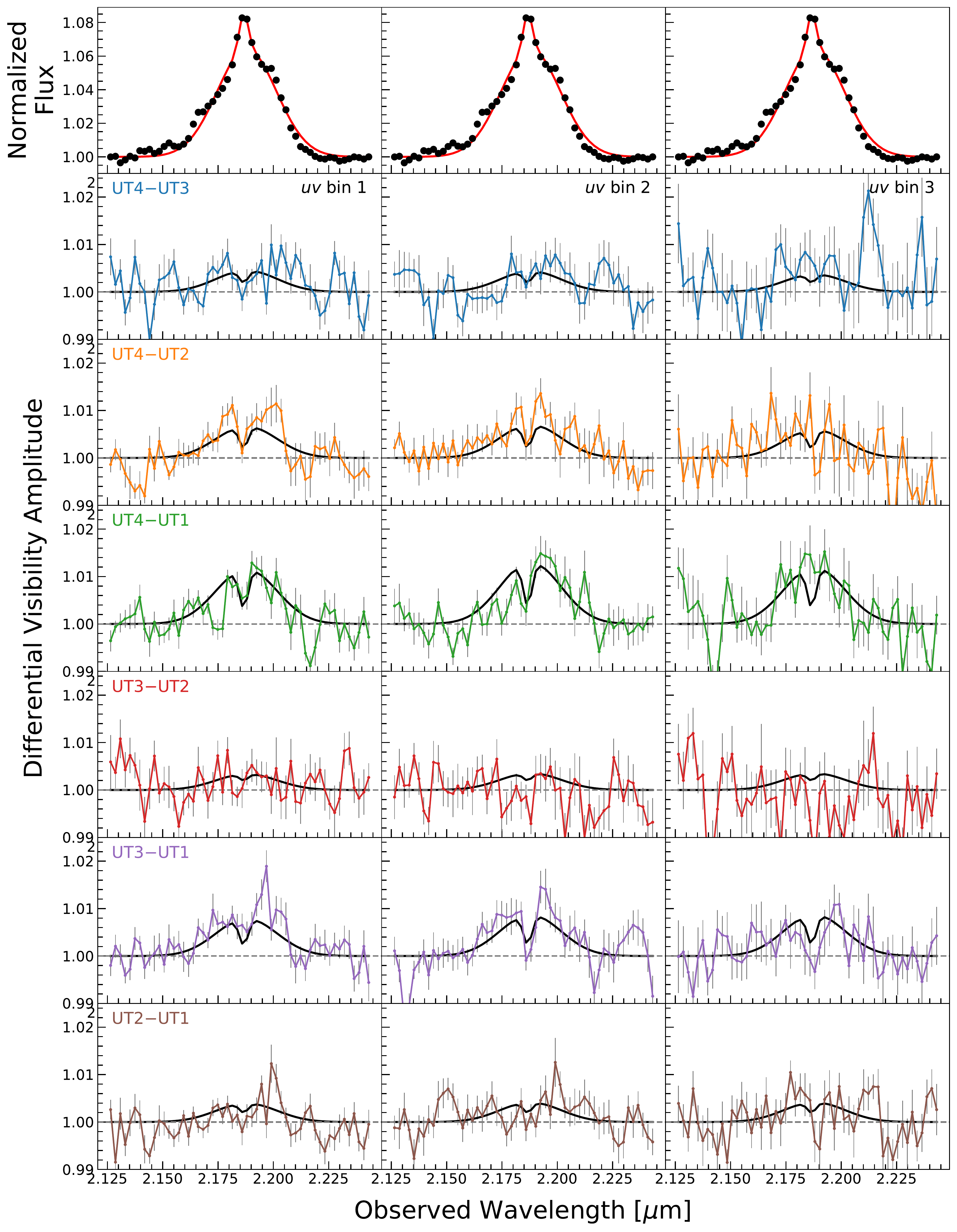}
    \caption{\textit{top row:} Normalized GRAVITY \brg{} profile (black points) with best fit model line profile (red line) from our coronal line region analysis. \textit{bottom rows:}$uv$-binned differential visibility amplitude spectra for each baseline (coloured points and lines) with the best fit CLR model (solid lines).}
    \label{fig:uvbin_bry_visamp}
\end{figure*}
\begin{figure*}
    \centering
    \includegraphics[width=0.8\textwidth]{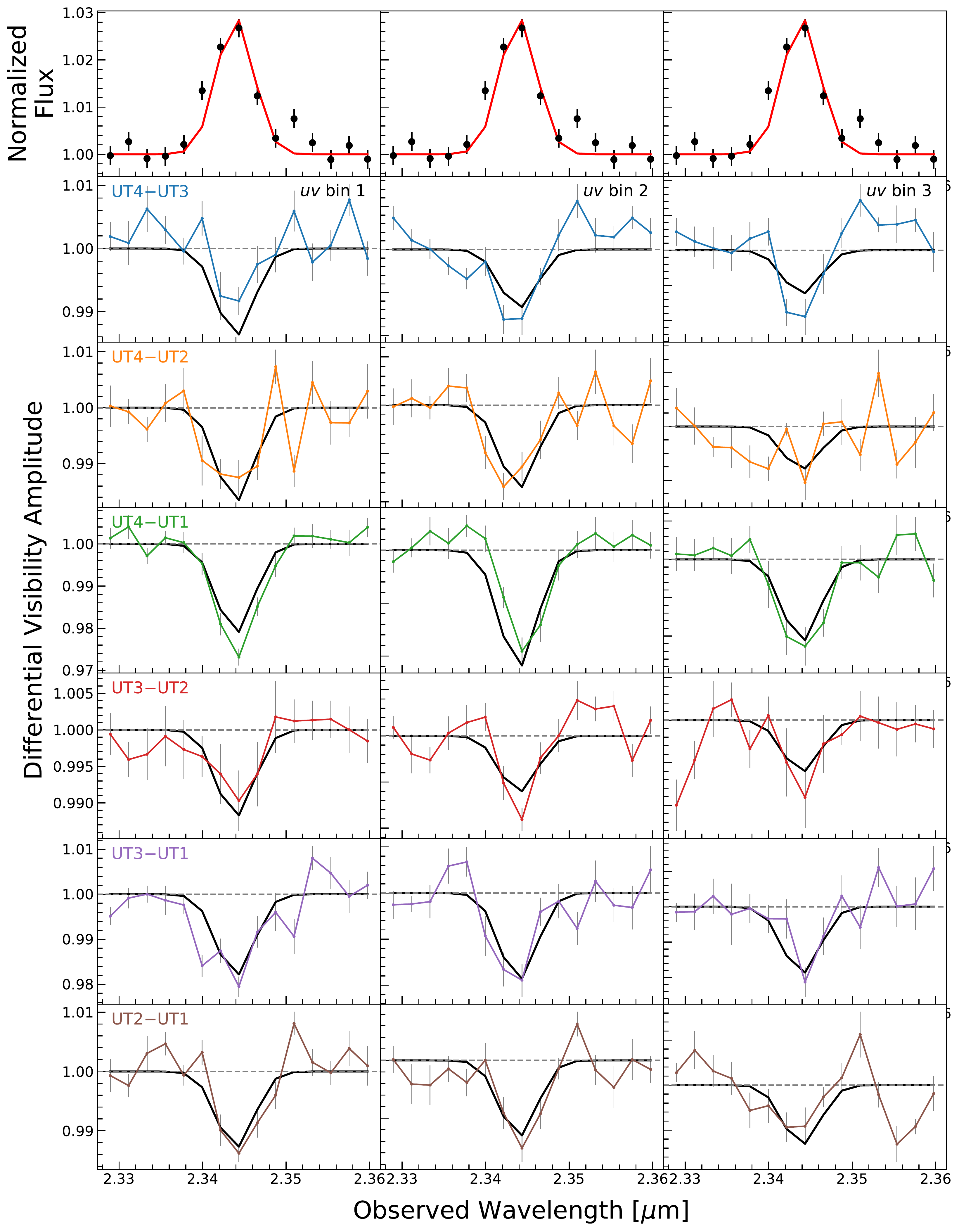}
    \caption{Same as Fig.~\ref{fig:uvbin_bry_visamp} but for the \caviii{} line.}
    \label{fig:uvbin_caviii_visamp}
\end{figure*}
\begin{figure*}
    \centering
    \includegraphics[width=\textwidth]{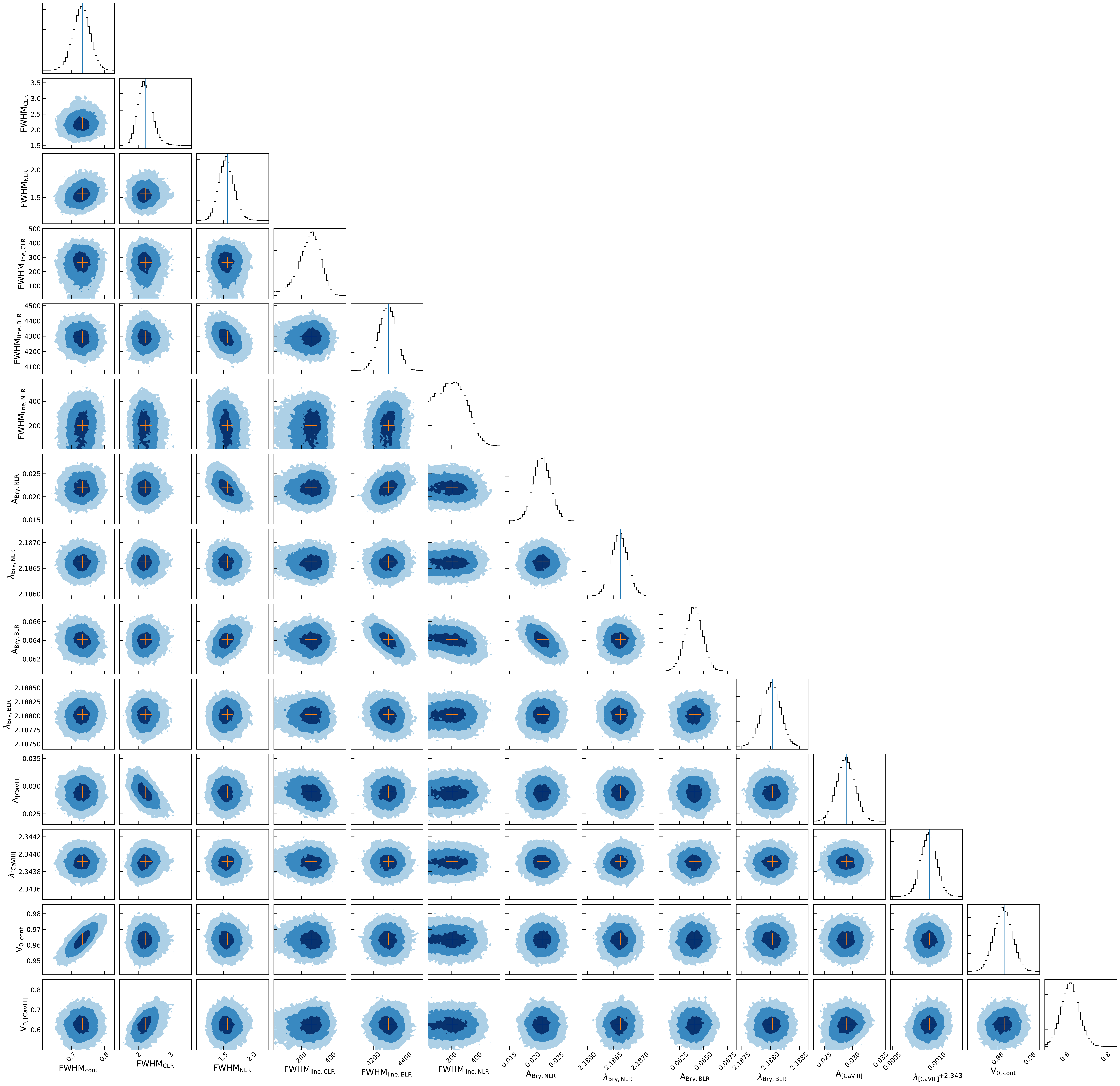}
    \caption{Corner plot for our CLR analysis showing the joint and marginalised posterior distributions for each free parameter. The blue lines and orange crosses indicate the maximum \textit{a posteriori} position used as the best fit values given in Table~\ref{tab:clr}}
    \label{fig:corner_clr}
\end{figure*}
\section{Image Reconstruction Robustness Tests}\label{app:mira_tests}
In this Appendix, we test the robustness of the features seen in our image reconstruction of the hot dust continuum for NGC~3783. We specifically tested the imaging as a function of 1) the choice of data, 2) the choice of regularisation, and 3) the choice of image reconstruction algorithm.

\subsection{Bootstrapping}

\begin{figure*}
    \centering
    \includegraphics[width=\textwidth]{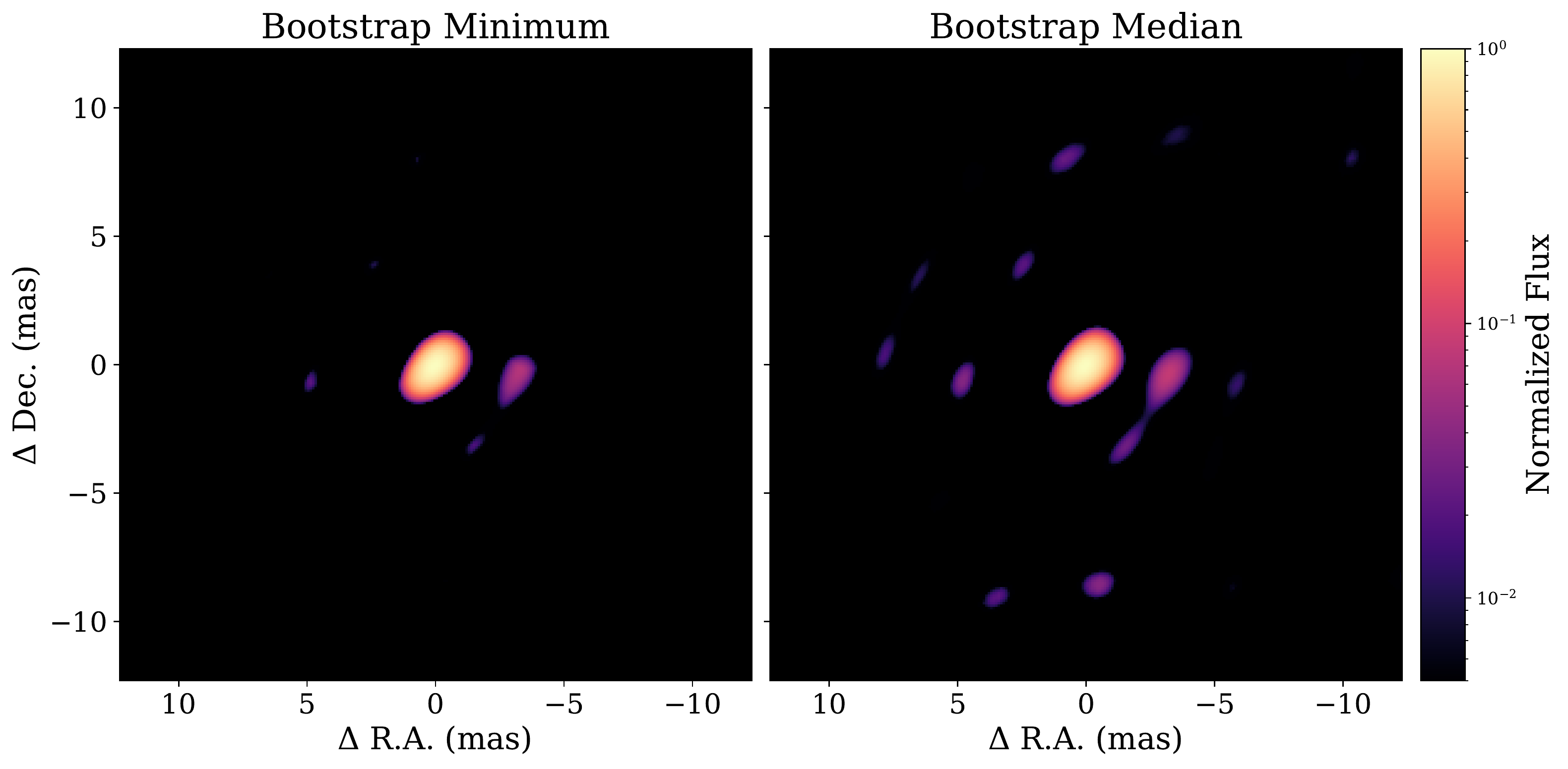}
    \caption{Minimum (left) and median images from our bootstrapping analysis of our image reconstruction. Values in the minimum image are determined as the minimum value from all 100 images produced. Values in the median image are the median value of all 100 images.}
    \label{fig:bootstrap_test}
\end{figure*}

Our first test involves testing the robustness of our image against which data are used in the reconstruction. We generated 100 sets of GRAVITY data with each set having 30\% of the original data randomly removed. We then ran \texttt{MiRA} on each of the 100 data sets to produce 100 images. From these 100 images, we calculated a minimum and median image. The minimum image is constructed using the minimum value of all 100 images for each pixel. The median is the median value of all 100 images for each pixel. These are shown in Fig.~\ref{fig:bootstrap_test}.

While the median image largely shows all the same features as the original image reconstruction, the minimum image instead has removed nearly all of the fainter features and left primarily the central source and the offset cloud. This strongly suggests that the fainter features are likely due to noise in the data and/or are artefacts of the image reconstruction. The offset cloud however is clearly a strong feature of the data that persists even after removing 30\% of the data.

\subsection{Regularisation}
\begin{figure}
    \centering
    \includegraphics[width=\columnwidth]{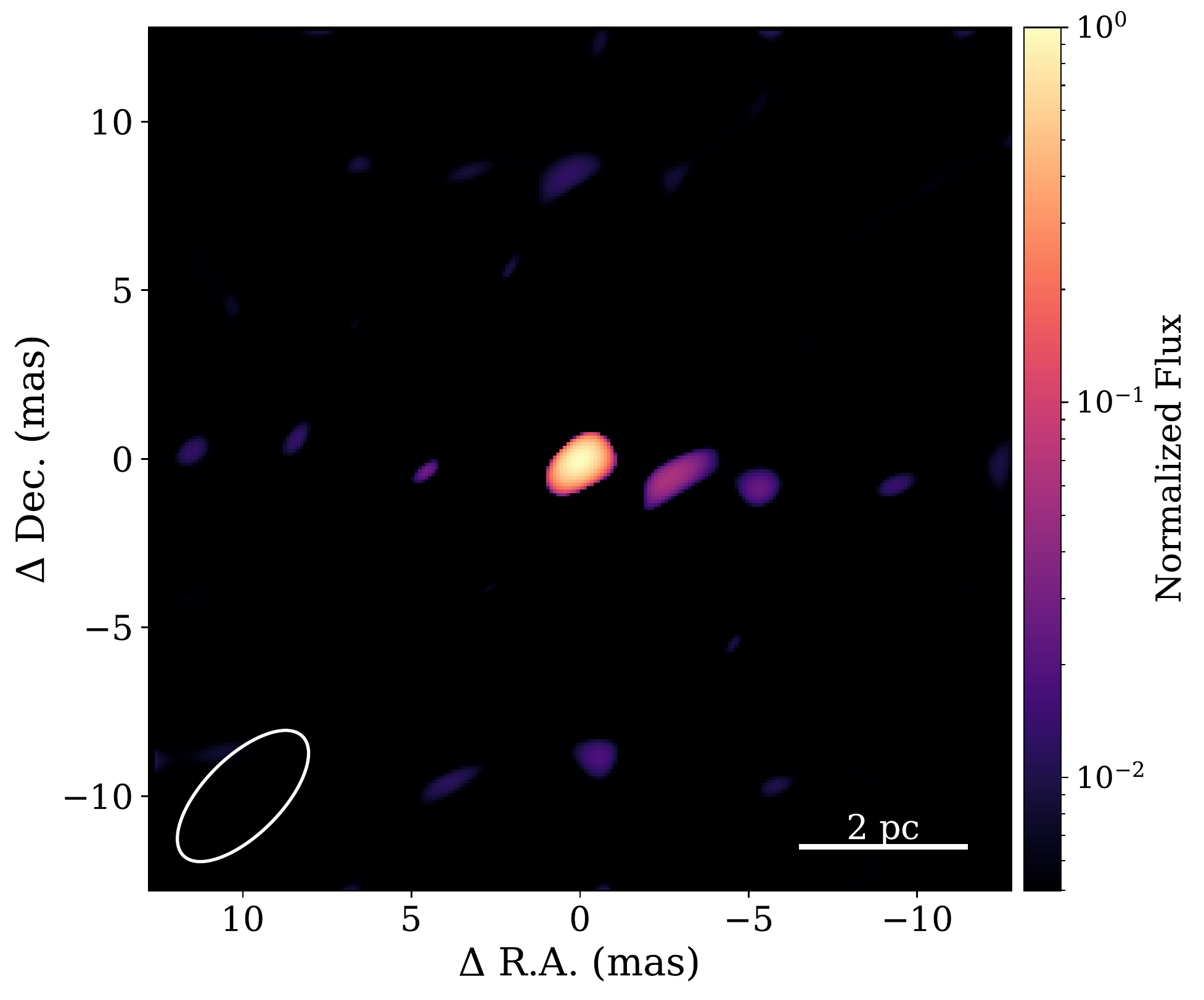}
    \caption{\texttt{MiRA} reconstructed image using the \textit{compactness} regularisation.}
    \label{fig:compactness_image}
\end{figure}

Our second test simply changes the specific regularisation used in the \texttt{MiRA} reconstruction. We instead applied the \textit{compactness} regularisation which prioritises centrally located compact sources. This is in contrast to the \texttt{hyperbolic} regularisation which prioritises smooth, extended sources with sharp edges. Fig.~\ref{fig:compactness_image} shows the reconstructed image under the \texttt{compactness} regularisation. 

As expected, the bright central source has decreased in size with a FWHM of 1.24x0.83 mas compared to 1.8x1.2 mas using the \texttt{hyperbolic} regularisation. However, the fainter offset cloud still remains at the same position relative the central source and contains the same 5\% of the total flux. This shows that the offset cloud is robust against our choice of regularisation available in \texttt{MiRA}.

\subsection{Image Reconstruction Algorithm}
\begin{figure}
    \centering
    \includegraphics[width=\columnwidth]{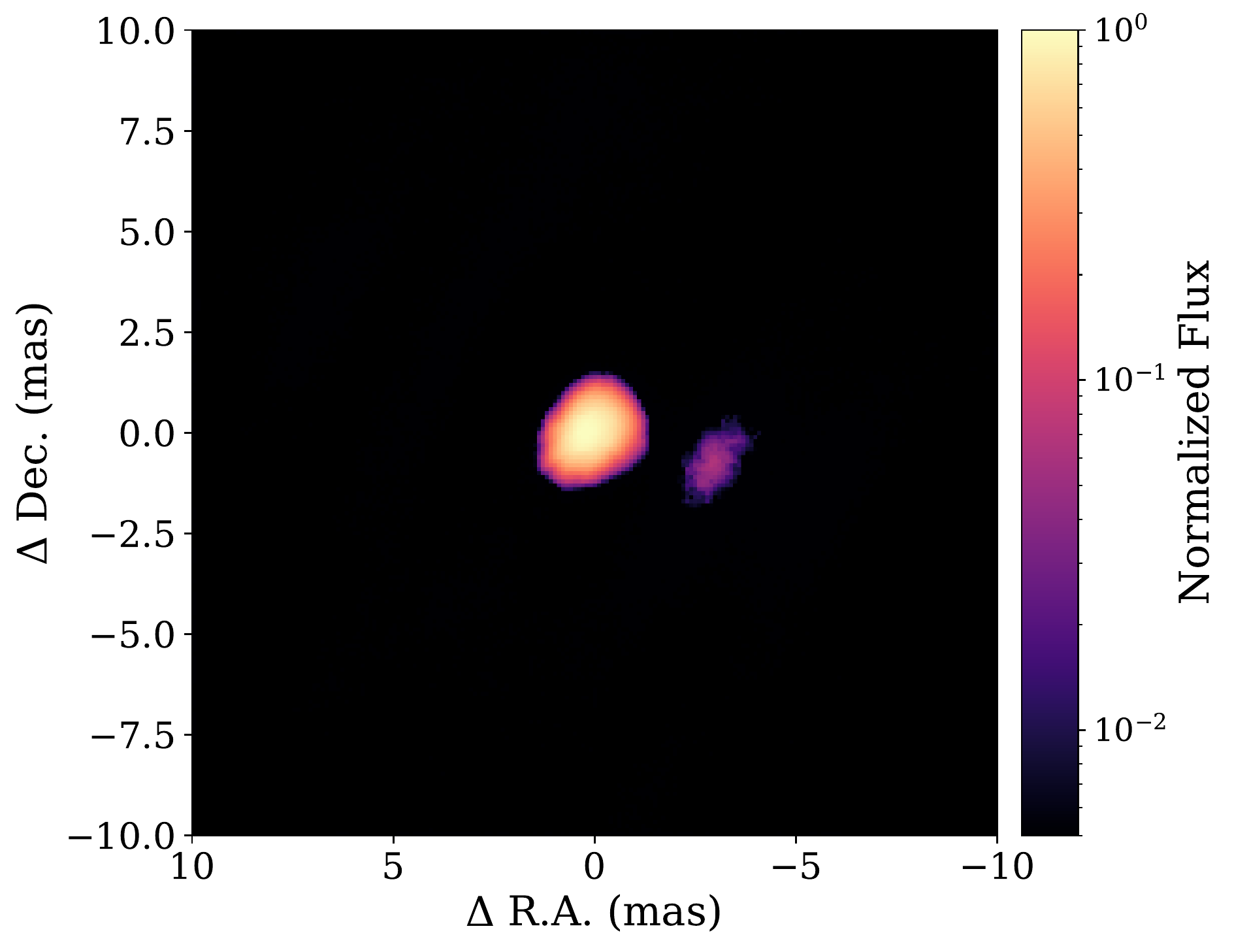}
    \caption{Image reconstruction of NGC~3783 using the \texttt{SQUEEZE} algorithm.}
    \label{fig:squeeze_image}
\end{figure}

For our final test, we used another image reconstruction algorithm, \texttt{SQUEEZE} \citep{Baron:2010aa} to ensure our image features are robust against the choice of image reconstruction algorithm. In Fig.~\ref{fig:squeeze_image}, we show the final median image reconstruction produced by \texttt{SQUEEZE}. Just as with \texttt{MiRA}, we see two components, a bright central extended component and an offset fainter component. The location and brightness of the offset component is very consistent with \texttt{MiRA} and our model fitting. Therefore, we conclude that the properties of the offset cloud is robust against our choice of image reconstruction algorithm.

\end{document}

%% file: tab1.tex
\begin{table}[]
\centering
\begin{tabular}{l|c|c|r@{$-$}l}
\hline\hline
Date        & On-source      & Seeing        & \multicolumn{2}{c}{Coherence} \\
            & time (min)     & (\arcsec)     & \multicolumn{2}{c}{time (ms)} \\ \hline
2018 Jan 07$^\ast$    & 80             & 0.38 $-$ 0.70 &  5.7 & 10.3 \\
2018 Jan 08$^\ast$ & 30             & 0.47 $-$ 0.58 &  7.8 & 10.6 \\
2018 May 31$^\ast$    & 95             & 0.38 $-$ 0.64 &  2.5 & 5.3 \\
2019 Feb 16$^\dagger$    & 140             & 0.45 $-$ 0.94 &  5.2 & 13.8 \\
2019 Mar 31$^\ast$    & 85             & 0.40 $-$ 0.67 &  3.0 & 4.9 \\
2020 Mar 08$^\dagger$           & 55             & 0.35 $-$ 0.78 & 4.6 & 10.0 \\
\hline\hline
\end{tabular}
\caption{Log of the VLTI/GRAVITY observations of NGC 3783 used in this work. On-source time in Column 2 reports only the frames for which the fringe tracking ratio was $>80$\%.  The seeing and coherence times are based on the measurements of the Differential Image Motion Monitor and Multi-aperture Scintillation Sensor on Paranal. \\
$^\ast$ K stars were observed as calibrators.  \\
$^\dagger$ B stars were observed as calibrators.}
\label{tab:obs}
\end{table}

%% file: tab2.tex
\begin{table}
\centering
\renewcommand{\arraystretch}{1.5}
%\begin{tabular}{l | c }
\begin{tabular}{p{0.45\columnwidth} | >{\centering\arraybackslash}p{0.45\columnwidth}}
\hline\hline
                          Parameters &        Best Fit Values                        \\ \hline 
$R_\mathrm{BLR}\,(\mu\mathrm{as})$   & $71_{-24}^{+56}$                             \\   
$R_\mathrm{min}\,(\mu\mathrm{as})$   & $32_{-18}^{+10}$                              \\
$\beta$                              & $1.7_{-0.9}^{+0.2}$                        \\ 
$\theta_\mathrm{o}\,(^\circ)$        & $24_{-11}^{+26}$                              \\ 
$i\,(^\circ)$                        & $23_{-10}^{+16}$                               \\ 
PA ($^\circ$ E of N)                 & $295_{-49}^{+55}$                             \\ 
$\kappa$                             & $-0.13_{-0.3}^{+0.5}$                                     \\            
$\gamma$                             & $1.5_{-0.4}^{+2.8}$                                       \\         
$\xi$                                & $0.7_{-0.6}^{+0.3}$                                       \\       
Offset ($\mu$as)                     & $\left(-0.5_{-16}^{+15}, -19_{-18}^{+11}\right)$ \\ 
$\log\,(\mbh/M_\odot)$               & $7.68_{-0.43}^{+0.45}$                         \\ 
$f_\mathrm{ellip}$                   & $0.46_{-0.38}^{+0.47}$                                             \\
$P(\mathrm{inflow})$                 & $0.65$                                       \\
$\theta_\mathrm{e}$                  & $45_{-42}^{+39}$                                       \\
%$\log\,\sigma_\mathrm{turb}$         & 0                                             \\ 
%$\lambda_\mathrm{emit}$              & $2.2892_{-0.0005}^{+0.0006}$                  \\
%                                     & $2.2923_{-0.0026}^{+0.0015}$       \\ \hline 
$\Delta v_\mathrm{BLR}\,(\mathrm{km\,s^{-1}})$  
                                     & $216_{-682}^{+150}$   \\ \hline   %& $380_{-356}^{+208}$      \\ \hline
$\chi_\mathrm{r}^2$                  & 0.665        \\           %& 1.38                     \\
\hline\hline
\end{tabular}
\caption{The best fit parameters and central 95\% credible interval for the modelling of the BLR spectrum
and differential phase of NGC~3783. Because $f_{\mathrm{flow}}$ is a binary switch, we instead report $P(\mathrm{inflow})$ which indicates the probability of inflow where $f_{\mathrm{flow}} < 0.5$. $\Delta v_\mathrm{BLR}$ is the difference between the velocity
derived from the best-fit $\lambda_\mathrm{emit}$ and the systemic velocity based on the redshift.
}
\label{tab:blr}
\end{table}

%% file: tab4.tex
\begin{table*}
\centering
\renewcommand{\arraystretch}{1.5}
\begin{tabular}{l | c c c c c c}
%\begin{tabular}{p{0.45\columnwidth} | >{\centering\arraybackslash}p{0.45\columnwidth}}
\hline\hline
Date        & $F_{\rm center}$ & FWHM$_{\rm center}$ & $F_{\rm offset}$  & x$_{\rm offset}$ & $y_{\rm offset}$ & $F_{\rm bkg}$ \\ 
\hline
2018 Jan 07 & $0.90 \pm 0.06$ & $0.73 \pm 0.03$      & $0.050 \pm 0.003$ & $-2.83 \pm 0.02$ & $-0.86 \pm 0.01$ & $0.05 \pm 0.01$ \\
2018 Jan 08 & $0.81 \pm 0.07$ & $0.79 \pm 0.04$      & $0.036 \pm 0.003$ & $-3.29 \pm 0.06$ & $-1.08 \pm 0.03$ & $0.16  \pm 0.02$  \\
2018 May 31 & $0.78 \pm 0.09$ & $0.66 \pm 0.04$      & $0.054 \pm 0.006$ & $-2.83 \pm 0.01$ & $-0.70 \pm 0.02$ & $0.17 \pm 0.02$ \\
2019 Feb 16 & $0.74 \pm 0.05$ & $0.60 \pm 0.03$      & $0.040 \pm 0.002$ & $-2.90 \pm 0.02$ & $-0.86 \pm 0.02$ & $0.22 \pm 0.01$ \\
2019 Mar 31 & $0.80 \pm 0.06$ & $0.68 \pm 0.03$      & $0.037 \pm 0.003$ & $-2.66 \pm 0.03$ & $-1.26 \pm 0.03$ & $0.17 \pm 0.01$ \\
2020 Mar 08 & $0.85 \pm 0.07$ & $0.88 \pm 0.02$      & $0.039 \pm 0.003$ & $-2.79 \pm 0.01$ & $-1.02 \pm 0.02$ & $0.11 \pm 0.01$ \\
\hline\hline
\end{tabular}
\caption{Best fit central Gaussian, offset point source, and background fits to the $V^2$ and closure phases for each night of observations. 
         $F_{\rm center}$, $F_{\rm offset}$, $F_{\rm bkg}$ are the fractional fluxes of each component. 
         FWHM$_{\rm center}$ is the Gaussian FWHM in mas of the Gaussian component and x$_{\rm offset}$ and $y_{\rm offset}$ are 
         the coordinates of the offset point source in mas relative to the Gaussian component. 
}
\label{tab:model_fits}
\end{table*}

%% file: tab3.tex
\begin{table}
\centering
\renewcommand{\arraystretch}{1.5}
%\begin{tabular}{l | c }
\begin{tabular}{p{0.45\columnwidth} | >{\centering\arraybackslash}p{0.45\columnwidth}}
\hline\hline
Parameters                                & Best Fit Values      \\ 
\hline
FWHM$_{\rm cont}$ (mas)                   & $0.73_{-0.05}^{+0.05}$    \\
FWHM$_{\rm [\ion{Ca}{viii}]}$  (mas)                   & $2.2_{-0.4}^{+0.5}$      \\
FWHM$_{\rm Br\gamma,n}$  (mas)                   & $1.6_{-0.2}^{+0.3}$      \\
FWHM$_{\rm line,[\ion{Ca}{viii}]}$ (km s$^{-1}$)       & $265_{-191}^{+123}$ \\
FWHM$_{\rm line,Br\gamma,b}$ (km s$^{-1}$)       & $4296_{-110}^{+109}$     \\
FWHM$_{\rm line,Br\gamma,n}$ (km s$^{-1}$)       & $202_{-181}^{+217}$ \\
$A_{\rm Br\gamma,n}$                    & $0.022_{-0.004}^{+0.004}$       \\
$v_{\rm Br\gamma,n}$ (km s$^{-1}$)       & $-79_{-41}^{+42}$    \\
$A_{\rm Br\gamma,b}$                    & $0.064_{-0.002}^{+0.002}$       \\
$v_{\rm Br\gamma,b}$ (km s$^{-1}$)       & $114_{-43}^{+41}$      \\ 
$A_{\rm [\ion{Ca}{viii}]}$                        & $0.03_{-0.003}^{+0.003}$       \\
$v_{\rm [\ion{Ca}{viii}]}$  (km s$^{-1}$)          & $30_{-22}^{+22}$     \\
$V_{0,\rm{Br\gamma}}$                     & $0.96_{-0.01}^{+0.01}$    \\
$V_{0,\rm{[\ion{Ca}{viii}]}}$                     & $0.63_{-0.09}^{+0.09}$    \\
\hline\hline
\end{tabular}
\caption{The best fit parameters and central 95\% credible interval 
for the modelling of the spectrum and differential visibility amplitude of NGC 3783.
$A_{\rm Br\gamma,n}$, $A_{\rm Br\gamma,b}$, $A_{\rm [CaVIII]}$ are the amplitudes of the narrow Br$\gamma$, broad Br$\gamma$, and \caviii{} lines and $v_{\rm Br\gamma,n}$, $v_{\rm Br\gamma,b}$, $v_{\rm [CaVIII]}$ are their central velocities relative to the redshift of NGC 3783.
}
\label{tab:clr}
\end{table}